\documentclass[manuscript,format=acmlarge,screen,nonacm]{acmart}

\usepackage{longtable}
\usepackage{lscape}
\usepackage{mdframed}
\usepackage{lipsum}
\usepackage{lscape}
\usepackage{tabularx}
\usepackage{multicol}
\usepackage{xcolor}
\usepackage{mathtools}

\usepackage{booktabs}
\usepackage{makecell}

\usepackage{pgfplots}
\pgfplotsset{compat=1.18}
\usepackage{pgfplotstable}
\usepackage{booktabs}
\usepackage{graphicx}
\usepackage{subcaption}
\usepackage{xcolor}
\usepackage{soul}

\newif\ifshowchanges

\ifshowchanges
    \includecomment{ifnotfinal} % Include content when showing changes
\else
    \excludecomment{ifnotfinal} % Exclude content when hiding changes
\fi

\makeatletter

\def\adl@drawiv#1#2#3{%
    \hskip.5\arrayrulewidth
    \leaders\hrule\@height#1\@depth#2\hskip#3
    \hskip.5\arrayrulewidth}
\makeatother

\let\hlBlue\hl
\renewcommand{\hl}[1]{{\sethlcolor{yellow}\hlBlue{#1}}}

\newcommand{\norm}[1]{\left\lVert#1\right\rVert}
\newcommand{\abs}[1]{|#1|}

\newcommand{\fmin}{200\,Hz}
\newcommand{\fmax}{1000\,Hz}
\newcommand{\ntaps}{256}
\newcommand{\delay}{64}
\newcommand{\StftWinsize}{512}
\newcommand{\StftStride}{64}
\newcommand{\rrWin}{20\,s}
\newcommand{\numPart}{18}

\newcommand{\numWinsVal}{2812}

\newcommand{\winDropped}{19.3\%}

\newcommand{\maeConf}{0.47}
\newcommand{\maeFuse}{0.84}
\newcommand{\maeSingle}{0.90}
\newcommand{\rmseConf}{1.65}
\newcommand{\rmseFuse}{2.26}
\newcommand{\rmseSingle}{2.35}

\newcommand{\maeFuseSD}{2.17}

\newcommand{\cpm}{\text{CPM}}

\newcommand{\sysName}{\text{EarResp-ANS}}

\newcommand{\Base}{\text{Baseline <35\,dB}}
\newcommand{\WNfifty}{\text{White Noise 50\,dB}}
\newcommand{\WNsixty}{\text{White Noise 65\,dB}}
\newcommand{\WNeighty}{\text{White Noise 80\,dB}}

\newcommand{\Music}{\text{Music}}

\newcommand{\tensilica}{\emph{Tensilica HiFi~3z}}
\newcommand{\nrf}{\emph{Nordic nRF5340}}

\newcommand{\RR}{\text{RR}}

\newcommand{\vect}[1]{\mathbf{#1}}

\newcommand{\RRhat}{\widehat{\mathrm{RR}}}

\newmdenv[linewidth=0pt,linecolor=white,innertopmargin=6,innerbottommargin=6,backgroundcolor=lightgray,skipbelow=8pt,skipabove=12pt,innerleftmargin=5,innerrightmargin=5,nobreak=true]{review}

\makeatletter
\def\@ACM@copyright@check@cc{}
\setcopyright{none}
\DeclareMathOperator*{\argmax}{arg\,max}

\newbool{showComments}
\booltrue{showComments}

\ifbool{showComments}{
\newcommand{\tr}[1]{{\sethlcolor{blue!20}\hlBlue{[Tobi R: #1]}}}

}

\begin{document}

%\input{sections/0-response-letter}

\title{\sysName{}: Audio-Based On-Device Respiration Rate Estimation on Earphones with Adaptive Noise Suppression}

\author{Michael K\"uttner}
\email{michael.kuettner@kit.edu}
\orcid{0009-0000-9021-0359}
\affiliation{%
  \institution{Karlsruhe Institute of Technology}
  \city{Karlsruhe}
  %\state{}
  \country{Germany}
}

\author{Valeria Zitz}
\email{valeria.zitz@kit.edu}
\orcid{0009-0004-1158-861X}
\affiliation{%
  \institution{Karlsruhe Institute of Technology}
  \city{Karlsruhe}
  %\state{}
  \country{Germany}
}

\author{Supraja Ramesh}
\email{supraja.ramesh@kit.edu}
\orcid{}
\affiliation{%
  \institution{Karlsruhe Institute of Technology}
  \city{Karlsruhe}
  %\state{}
  \country{Germany}
}

\author{Michael Beigl}
\email{michael.beigl@kit.edu}
\orcid{0000-0001-5009-2327}
\affiliation{%
  \institution{Karlsruhe Institute of Technology}
  \city{Karlsruhe}
  %\state{}
  \country{Germany}
}

\author{Tobias R\"oddiger}
\email{tobias.roeddiger@kit.edu}
\orcid{0000-0002-4718-9280}
\affiliation{%
  \institution{Karlsruhe Institute of Technology}
  \city{Karlsruhe}
  %\state{}
  \country{Germany}
}

\renewcommand{\shortauthors}{K\"{u}ttner et al.}

% 150 - 200 words
\begin{abstract}
Respiratory rate (\RR{}) is a key vital sign for clinical assessment and mental well-being, yet it is rarely monitored in everyday life due to the lack of unobtrusive sensing technologies. In-ear audio sensing is promising due to its high social acceptance and the amplification of physiological sounds caused by the occlusion effect; however, existing approaches often fail under real-world noise or rely on computationally expensive models.
We present \sysName{}, the first system enabling fully on-device, real-time \RR{} estimation on commercial earphones. The system employs LMS-based adaptive noise suppression (ANS) to attenuate ambient noise while preserving respiration-related acoustic components, without requiring neural networks or audio streaming, thereby explicitly addressing the energy and privacy constraints of wearable devices.
We evaluate \sysName{} in a study with 18 participants under realistic acoustic conditions, including music, cafeteria noise, and white noise up to 80\,dB SPL. \sysName{} achieves robust performance with a global MAE of \maeFuse{}\,\cpm{}, reduced to \maeConf{}\,\cpm{} via automatic outlier rejection, while operating with less than 2\% processor load directly on the earphone.
\end{abstract}




%     \centering
%      \includegraphics[width=\linewidth,page=1,trim=0cm 9.5cm 0cm 0cm]{}
%     \caption{}
%     \label{fig:teaser-figure}

\maketitle

\section{Introduction}
\label{sec:intro}

% --- 1. Clinical Significance and the Monitoring Gap ---
Respiration rate (\RR{}) is a key vital sign providing critical insights into a person’s physiological~\cite{loughlin2018respiratory} and mental state~\cite{grossman1983respiration}. It plays a vital role in sleep monitoring~\cite{gutierrez2016respiratory}, stress estimation~\cite{hernando2015changes}, and the assessment of physical fitness~\cite{nicolo_importance_2020, vitazkova_advances_2024}. Furthermore, \RR{} is often one of the earliest indicators of physiological deterioration~\cite{loughlin2018respiratory}, making it highly relevant for detecting emerging health issues such as chronic obstructive pulmonary disease~\cite{shah2017exacerbations} or sepsis~\cite{yang2025early}, which are significant drivers of mortality~\cite{halpin2006chronic, winters2010long}. Despite its clinical importance, \RR{} is still frequently under-monitored~\cite{kayser2023respiratory, cretikos2008respiratory, loughlin2018respiratory} or inaccurately captured in practice~\cite{drummond2020current, kallioinen2021quantitative}.

In clinical settings, manual counting is error-prone~\cite{kallioinen2021quantitative}. Professional medical devices typically rely on more robust techniques like spirometry~\cite{american2019standardization} or respiratory inductance plethysmography (RIP)~\cite{carry1997evaluation}, which are often bulky, socially unacceptable, and unsuitable for continuous everyday wear. 
More portable wearable approaches, such as those based on inertial measurement units (IMUs), are more direct but highly susceptible to motion artifacts~\cite{roddiger2019towards, hernandez2015cardiac, hernandez2015biowatch}.
Photoplethysmography (PPG)-based approaches are less sensitive to motion but typically provide only indirect estimation metrics~\cite{meredith2012photoplethysmographic}.
%More portable wearable approaches, such as those based on photoplethysmography (PPG), often only provide indirect estimation metrics~\cite{meredith2012photoplethysmographic}. Inertial measurement unit (IMU)-based approaches are more direct but highly susceptible to motion artifacts~\cite{roddiger2019towards, hernandez2015cardiac, hernandez2015biowatch}.

% --- 2. The Potential of Microphone-Based Earables ---
Microphone-based sensing provides a promising alternative, as breathing sounds directly encode respiratory activity~\cite{romano2023respiratory, hu2024breathpro, liu2025respear}. Earphones are particularly promising as they allow continuous sensing in close vicinity to the airways and are increasingly viewed as versatile sensing platforms for health monitoring~\cite{CLARKE2025207}. 
A key advantage is that earphones are a socially accepted class of devices routinely worn at work, while commuting, or during sports~\cite{harshitha2017survey}. 
Beyond active playback, earphones are frequently worn to create a private auditory space. 
Empirical studies show that 57.2\% of users wear personal listening devices primarily to block external noise, often without audio playback~\cite{haas2018can}, which can reduce sensory stress ("Walkman effect")~\cite{hosokawa1984walkman, waldecker2017ohren, radun2024active}. 
This makes earphones a particularly suitable platform for unobtrusive physiological sensing. 
Integrating respiration tracking into earphones enables multipurpose health monitoring (e.g., \RR{}-based stress management) without additional, stigmatizing medical equipment~\cite{grossman1983respiration, CLARKE2025207}.
%A key advantage is that earphones are a socially accepted class of devices routinely worn at work, while commuting, or during sports~\cite{harshitha2017survey}. Beyond active playback, earphones are frequently worn for two everyday reasons. First, they support \emph{behavioral soundscape curation}: 57.2\% of users wear personal listening devices primarily to block external noise, often without audio playback~\cite{haas2018can}, which can reduce sensory stress ("Walkman effect")~\cite{hosokawa1984walkman, waldecker2017ohren, radun2024active}. Second, integrating \sysName{} enables unobtrusive, multipurpose health monitoring (e.g., \RR{}-based stress management) without additional, stigmatizing equipment~\cite{grossman1983respiration, CLARKE2025207}.

In contrast to specialized body-worn sensors with visible straps~\cite{harshitha2017survey}, in-ear microphones (IEMs) used in noise-canceling earphones benefit from the occlusion effect~\cite{OESense2021}. The occlusion effect amplifies internally generated sounds such as breathing~\cite{BreathPro2024, martin2017ear}, thereby improving the signal-to-noise ratio (SNR) compared to externally worn microphones.
% --- 3. Technical Challenges ---
However, making microphone-based sensing feasible on commodity earphones remains challenging:
\begin{itemize}
    \item \textbf{Acoustic Interference:} Low-amplitude breathing sounds are easily masked by speech and environmental noise~\cite{romano2023respiratory, kumar2021estimating, xu2019breathlistener}.
    \item \textbf{Resource Constraints:} Continuous audio processing is computationally demanding for platforms with limited compute and strict energy budgets~\cite{roddiger2025openearable}.
    \item \textbf{Privacy and Latency:} Offloading raw audio raises privacy concerns and increases energy consumption and latency~\cite{BreathPro2024, martin2017ear}.
\end{itemize}

% --- 4. The Proposed Solution ---
To address these challenges, we present \textbf{\sysName{}}, a fully on-device \RR{} estimation pipeline. We exploit the dual microphone configuration (IEM/OEM) to suppress environmental noise prior to inference. Using an adaptive noise suppression (ANS) approach based on a Least Mean Squares (LMS) adapted filter, we eliminate external noise while preserving the respiratory features amplified by the occlusion effect. \sysName{} estimates \RR{} via lightweight spectral analysis, avoiding neural inference and raw audio streaming. Furthermore, it leverages binaural fusion from both earphones to detect unreliable estimates.

% --- 5. Contributions ---
We evaluate \sysName{} in a study with \numPart{} participants under realistic conditions (80\,dB music, cafeteria noise) and deploy our system on the OpenEarable~2.0 platform~\cite{roddiger2025openearable}. We demonstrate that:
\begin{enumerate}
    \item \textbf{\sysName{}} provides a fully on-device, real-time \RR{} estimation pipeline using resource-efficient spectral analysis, imposing less than 2\% computational load on the earphones' processors.
    \item Our \textbf{dual-microphone ANS strategy} and binaural consistency check substantially improve robustness under ambient noise of up to more than 80\,dB sound pressure level (SPL) by preserving respiration-related signal components.
    \item The system establishes a \textbf{new state-of-the-art} for noise-robust on-device respiration rate sensing on earphones~\cite{roddiger2025openearable}, achieving a mean absolute error (MAE) of \maeConf{} cycles per minute (\cpm{}) after automatic outlier removal, surpassing prior audio-based approaches.
\end{enumerate}

\section{Related Work}
\label{sec:related-work}
%\tr{reduce to three pages}

This section reviews prior work on wearable respiration rate monitoring, focusing on approaches that enable continuous tracking using wearables and form factors suitable for everyday use (see \autoref{subsec:wearable-respiration}). In addition, we review prior work on noise reduction and signal enhancement for wearable audio sensing, with an emphasis on methods applicable to real-time, on-device processing (see \autoref{sec:rt_ns}).

%We organize the literature by sensing modality, covering inertial-based methods, heart-rate-based approaches that exploit cardiovascular modulation, and audio-based systems that estimate respiration from breathing sounds. 
%Across these categories, we discuss differences in signal directness, robustness to motion and environmental interference, evaluation settings, and system-level constraints such as computational load, energy consumption, and privacy. 

\subsection{Wearable Respiration Monitoring Systems}
\label{subsec:wearable-respiration}
Wearable respiration monitoring enables continuous assessment of breathing under everyday conditions, where environmental noise and acoustic interference pose significant challenges to signal reliability~\cite{jacobsen_noninvasive_2021,hsueh_comprehensive_2025}.
Existing approaches can be broadly grouped by sensing modality into inertial-based methods, heart-rate-based methods exploiting cardiovascular modulation, and audio-based methods that capture breathing sounds directly. These modalities differ in signal directness, robustness to interference, and suitability for continuous, on-device operation. \autoref{tab:survey} summarizes representative systems across these categories.
% wordy and not helpful, removed. Rather than serving as a quantitative benchmark, the table highlights the diversity of sensing assumptions, evaluation protocols, and deployment constraints that shape current wearable respiration monitoring approaches.

\newcolumntype{L}[1]{>{\raggedright\arraybackslash}p{#1}}
\newcolumntype{Y}{>{\raggedright\arraybackslash}X}

\begin{table}[!t]
\footnotesize
\setlength{\tabcolsep}{3.5pt}
\renewcommand{\arraystretch}{1.15}
\caption{Overview of representative respiration rate monitoring approaches. Motion and modulation categories reflect the experimental conditions reported in the respective studies. Reported error metrics are not directly comparable due to differences in reference measurements, evaluation protocols, and motion conditions across studies.}
\centering
\begin{tabularx}{\textwidth}{
l                  % Work 
l                  % Sensor
Y                  % Placement
l                  % Motion
l                  % Modulation
l                  % Method
Y                  % MAE
l                  % SD
l                  % Deployed
@{}}
\toprule
\textbf{Work} &
\textbf{Sensor} &
\textbf{Placement} &
\textbf{Motion} & %Explicit: real-world / walking / daily activities %None: sedentary, no motion analysis; Limited: leichte Bewegung / implizit; Partial: Bewegung explizit erwähnt, aber nicht frei
\textbf{Modulation} &
\textbf{Method} & % ML or SP or SP + ML
\textbf{MAE} &
\textbf{SD} &
\textbf{Deployed} \\ %On-device or Edge device (Phone, Computer) or -
\midrule

Angelucci et al.~\cite{angelucci_imu-based_2023} &
IMU &
Multi &
Resting, Motion &
None, Exercise &
SP + ML &
-- &
-- &
-- %
\\

%BioWatch~\cite{hernandez2015biowatch} &
%IMU &
%Wrist &
%Resting &
%None &
%SP & %filters, FFT
%0.38 &
%1.19 & %just gyroscope, acc. + gyr.:0.55 1.80
%-- %
%\\

R\"oddiger et al.~\cite{roddiger2019towards} &
IMU &
In-ear &
Resting &
None &
SP &
2.62 &
2.74 &
-- % info fehlt
\\

OptiBreathe~\cite{romero2024optibreathe} &
PPG &
In-ear &
Resting &
None, Instructed &
SP &
$\approx2.7^{*}$ / $1.96^{\dagger}$ &
-- & %0.2 for 15bpm instructed breathing, \approx 0.27 for no modulation (natural breathing)
-- % info fehlt
\\

Romano et al.~\cite{romano2023respiratory} &
Mic. &
Face &
Resting, Motion &
None, Exercise &
SP &
1.7 &
-- &
-- %third-order Butterworth bandpass filter, Hilbert transform, moving average filter, Butterworth bandpass filter as well as Ambient Noise Estimation with calculated the sound pressure level sing the microphone positioned outside the facemask --> nicht als Eingangsgröße benutzt, um die Respiratory Rate Estimation (fR-Schätzung) zu korrigieren oder zu verbessern, wurde separat berechnet und dann zur Auswertung/Interpretation der Messgüte herangezogen.
\\

RespEar~\cite{liu2025respear} &
Mic. &
In-ear &
Resting, Motion &
None, Exercise &
SP + ML &
$1.48^{\ddagger}$ / $2.28^{\S}$ &
-- & %Multirow: Motion: 2.28
Phone
\\

Kumar et al.~\cite{kumar2021estimating} &
Mic. &
In-ear &
Resting, Motion &
None, Exercise &
ML &
-- &
-- &
-- %
\\

Martin et al.~\cite{martin2017ear} &
Mic. &
In-ear &
Resting &
None &
SP &
2.7 &
1.6 &
--
\\

BreathPro~\cite{BreathPro2024} &
Mic. &
In-ear &
Motion &
Exercise &
SP + ML &
-- &
-- &
Phone %The custom Raspberry Pi–based prototype was used to collect synchronized in-ear and out-ear audio data during outdoor running. The proposed signal processing and machine learning pipeline was evaluated offline.
\\

%\midrule
\textbf{\sysName{} (ours)} &
\textbf{Mic. }&
\textbf{In-ear} &
\textbf{Resting} &
\textbf{None, Exercise} &
\textbf{SP }&
\textbf{\maeFuse{} }&
\textbf{\maeFuseSD{}} &
\textbf{Earphones}
\\

\bottomrule
\end{tabularx}

\vspace{2pt}
\footnotesize
\emph{Notes:}
$^{*}$ No modulation;
$^{\dagger}$ Instructed breathing at 15 bpm;
$^{\ddagger}$ Resting condition;
$^{\S}$ Motion condition.
\label{tab:survey}
\end{table}

\paragraph{Inertial-Based Respiration Rate Estimation}
Inertial approaches estimate respiration rate from breathing-induced body motion captured by accelerometers and gyroscopes~\cite{rodrigo2024}. 
Because IMUs are low-power and already integrated in many wearables, they are widely used for respiration monitoring.
%Most systems place sensors on the thorax or abdomen, where respiratory motion is strongest~\cite{yoon2014improvement,de2019use,chang2020portable}. A recent review reports typical MAE on the order of 0.5--2~\cpm\ under controlled conditions (often rest or sleep)~\cite{rodrigo2024}. 
To improve wearability, different placements have been explored, including head-mounted sensing~\cite{hernandez2015cardiac}, %wrist-worn devices~\cite{hernandez2015biowatch}, 
and ear-worn IMUs~\cite{roddiger2019towards}.  
%While accurate under controlled conditions, posture changes and gross motion introduce artifacts that overlap with the respiratory frequency band, limiting robustness in daily-life scenarios~\cite{rodrigo2024, roddiger2019towards}. Although some systems mitigate this issue by discarding motion-affected windows and estimating respiration primarily during rest~\cite{roddiger2019towards}, this problem is maximized during motion~\cite{rodrigo2024}.
While accurate under controlled conditions, posture changes and gross motion introduce artifacts that overlap with the respiratory frequency band, limiting robustness in daily-life scenarios~\cite{rodrigo2024, roddiger2019towards}.
Multi-sensor setups and activity-aware processing can improve robustness but increase system complexity and energy consumption, which is at odds with lightweight continuous monitoring~\cite{angelucci_imu-based_2023}. 
%wordy, removed. Because respiratory motion occupies the same low-frequency band as many everyday movements, this limitation is fundamental rather than algorithmic, and cannot be fully resolved by filtering or learning-based separation.

\paragraph{Heart-Rate-Based Respiration Rate Estimation}

Heart-based approaches infer respiration rate indirectly from respiration-induced modulations of cardiovascular signals measured by Electrocardiography (ECG) or PPG~\cite{charlton2017extraction}. 
Physiologically, breathing induces changes in inter-beat intervals (respiratory sinus arrhythmia)~\cite{yasuma2004respiratory} and causes amplitude/baseline variations in PPG waveforms~\cite{charlton2017extraction,karlen2011respiratory}. 
Based on these effects, respiration can be estimated using classical spectral and time-domain methods under low-motion conditions, and more recent work explores multimodal fusion (e.g., ECG+PPG) to increase robustness~\cite{john_multimodal_2025}.
Although effective under low-motion conditions, these methods degrade under motion, changing contact pressure, peripheral perfusion variations, and ambient light interference, limiting their suitability for continuous wearable monitoring~\cite{charlton2017extraction,prabakaran2022review}.

\paragraph{Audio-Based Respiration Rate Estimation}
Audio-based approaches estimate respiration rate directly from breathing sounds captured by microphones, providing a more direct sensing modality compared to inertial or cardiovascular proxies~\cite{romano2023respiratory,kumar2021estimating}. Early work demonstrated feasibility using classical signal processing techniques under controlled acoustic conditions but did not systematically address robustness to non-stationary environmental noise or overlap with speech and ambient sounds~\cite{martin2017ear,romano2023respiratory}. This limitation motivates the need for explicit background noise suppression, reviewed in \autoref{sec:rt_ns}.

Prior work has explored different microphone placements, including wearable microphones near the airway and head-mounted microphones, to capture breathing sounds during rest and moderate activity~\cite{kumar2021estimating,romano2023respiratory}. 
However, breath sounds are often low in amplitude and are easily masked by environmental noise, ambient acoustic interference, and motion-induced artifacts, resulting in substantial performance degradation under realistic everyday conditions~\cite{romano2023respiratory,xu2019breathlistener}. 
%While these studies established feasibility, /citet{martin2017ear} largely assumed constrained acoustic environments and did not explicitly investigate robustness against non-stationary environmental noise typical of everyday wearable use. %In particular, prior spectral approaches did not examine how respiration-related acoustic cues can be reliably isolated when breathing sounds overlap with speech or ambient noise. We therefore review prior work on real-time microphone noise reduction and earable-specific denoising strategies in \autoref{sec:rt_ns}.

%Building on this foundation, recent work has investigated earable platforms equipped with in-ear microphones as a socially acceptable and unobtrusive form factor for long-term physiological sensing~\cite{martin2017ear,liu2025respear,BreathPro2024}. 
Recent work has therefore increasingly explored earable platforms as a socially acceptable and unobtrusive form factor for long-term audio-based physiological sensing. In-ear microphones
%\mk{die in-ear micrphones kommen mir etwas aus dem nichts, es fehlt vielleicht die Verbindung zu earphones} 
benefit from the occlusion effect, which amplifies internally generated physiological sounds such as breathing within the sealed ear canal, and earphones are already widely adopted in everyday life, making them an attractive platform for continuous respiration monitoring~\cite{OESense2021}. While the occlusion effect amplifies internally generated physiological sounds such as breathing, in-ear microphone recordings remain susceptible to environmental noise due to acoustic leakage through the earphone housing, bone-conduction pathways, and imperfect sealing~\cite{CLARKE2025207,roddiger2025openearable,OESense2021,BreathPro2024}. %\mk{das hatten wir schonmal oder :D}.
%As a result, respiration signals recorded in realistic everyday environments are often strongly corrupted by artifacts induced by noise, speech etc., leading to substantial performance degradation.

To address these challenges, several recent systems rely on learning-based models to separate breathing sounds from background noise and other acoustic events to measure respiration volume~\cite{EarMeter2025} or reconstructing a fine-grained breathing waveform (from which respiration rate can be derived) in noisy driving~\cite{xu2019breathlistener}. However, such approaches typically require large labeled datasets and substantial computational resources, limiting their suitability for continuous monitoring on resource-constrained wearables and raising concerns about energy consumption, latency, and user privacy, especially when raw audio is processed off-device~\cite{zhang_survey_2025,CLARKE2025207}.
In contrast, systems such as BreathPro~\cite{hu2024breathpro} rely on signal-processing–based noise reduction without learning-based denoising, while using lightweight machine learning only for downstream respiration-rate estimation~\cite{BreathPro2024}.
These limitations motivate a lightweight, signal-processing–based audio approach for respiration-rate estimation that is robust to environmental noise and evaluated directly on earphones for continuous, on-device operation in terms of energy and latency; our approach follows this direction.

\subsection{Real-time Audio Background Noise Suppression}\label{sec:rt_ns} 
Background noise such as speech, traffic, or ambient environmental sounds significantly degrades microphone recordings in everyday settings. Because breathing sounds are low-amplitude, broadband, and highly variable, they often exhibit very low SNR in realistic conditions~\cite{owino2025}, making audio-based respiration-rate estimation unreliable without effective noise suppression. For wearable deployment, such methods must operate in real time, run fully on-device under strict computational and energy constraints, and avoid transmitting raw audio to preserve user privacy. These requirements motivate a closer examination of background noise suppression techniques suitable for earable-based respiration sensing.

\paragraph{Audio-Based Noise Reduction: General Single-Channel Approaches} %\mk{context based?}
In single-channel noise reduction, the target signal and interference are fully mixed within one microphone recording, rendering source separation inherently ill-posed due to the absence of spatial, causal, or reference information~\cite{Widrow1975ANC}.

%As a result, noise suppression is performed through inference rather than physical separation, relying on assumptions about the statistical, spectral, or temporal characteristics of the involved signals. 
Classical model-based approaches, including Spectral Subtraction~\cite{Boll1979}, Wiener Filtering~\cite{Modhave2016hearingaids,purushotham2018adaptive}, and minimum mean square error (MMSE) estimation~\cite{Ephraim1984}, exploit assumptions such as stationarity, spectral sparsity, or energy dominance to attenuate unwanted components. While these methods offer low computational complexity and predictable latency, their effectiveness strongly depends on the validity of the underlying assumptions, which are frequently violated in dynamic real-world environments~\cite{UPADHYAY2015574}.

Recent deep learning–based noise reduction methods replace explicit signal models with data-driven mappings learned from large-scale datasets, enabling improved suppression of non-stationary interference in many scenarios~\cite{yousif2025speech}. However, these approaches remain fundamentally constrained by the same single-channel observation: separation is achieved by learning statistical correlations rather than by exploiting physical source separation cues. Consequently, model behavior is tightly coupled to the characteristics of the training data and evaluation objectives, leading to reduced robustness under domain shift and unseen acoustic conditions~\cite{Lee2019, lin2023two}. This limitation has been consistently observed across bioacoustic and audio sensing domains, where high performance under controlled benchmarks often fails to translate to real-world deployments~\cite{owino2025}.

For weak, highly variable target signals, such as physiological sounds, these limitations become particularly pronounced. In the absence of explicit reference channels, single-channel noise reduction methods tend to either suppress relevant signal components or introduce temporal smearing and instability, especially under low signal-to-noise ratios and strict real-time constraints~\cite{Jingdong2006}. Importantly, these effects arise not from specific algorithmic choices but from the fundamental ambiguity of the single-channel formulation itself.

\paragraph{Audio-Based Noise Reduction: Reference-Channel–Based Approaches} %\mk{reference channel based?} 

Reference-channel–based noise reduction extends single-channel formulations by incorporating an additional microphone that captures interference correlated with the target signal, enabling physically informed noise suppression via cross-channel correlation~\cite{Widrow1975ANC}. Classic adaptive filtering techniques such as LMS-based approaches learn a time-varying mapping from the reference channel to the interference component under real-time constraints, avoiding the fundamental ambiguity of single-channel separation and improving robustness at low signal-to-noise ratios~\cite{Widrow1975ANC,haykin2001minimum}.

In earphone and earable systems, the physical separation between an outer-ear microphone (OEM) and an in-ear microphone (IEM) naturally provides a reference–target configuration: the OEM primarily captures ambient acoustic interference, while the IEM records a mixture of externally coupled noise and internally generated physiological sounds modulated by the occluded ear canal~\cite{bouserhal2017ear}. Prior work has explored adaptive filtering in this configuration for physiological sensing; Martin et al.~\cite{martin2017ear} demonstrate robust heart-rate estimation under artificially added industrial noise but substantially degraded breathing-rate performance, highlighting the need to preserve respiration-specific information.

More recently, BreathPro~\cite{BreathPro2024} leverages an out-ear microphone as a noise reference for in-ear breathing sensing during running, but relies on precomputed, user-specific templates rather than continuous online adaptation, leading to performance degradation when generalized templates are used~\cite{BreathPro2024}. While active noise cancellation (ANC) systems are traditionally designed for perceptual noise reduction, their feedforward reference-channel and adaptive filtering principles are directly applicable to numerical noise suppression for audio sensing~\cite{kuo2002active}. However, existing approaches do not explicitly address adaptive, signal-preserving noise suppression for continuous, fully on-device respiration rate estimation on earphones under realistic and changing acoustic conditions; a technical discussion of ANC principles and reference-channel adaptive filtering is provided in \autoref{sec:background}.

%Building on these observations, this work revisits classical reference-channel adaptive filtering in the context of in-ear respiration sensing. Rather than introducing new denoising models, we investigate how lightweight LMS-based adaptive noise suppression, that is traditionally used for ANC and speech enhancement, can be repurposed to preserve respiration-specific temporal structure under realistic environmental noise. To this end, we design \sysName{}, a fully on-device system that integrates adaptive noise suppression, respiration rate estimation, and channel-discrepancy–based outlier handling into a single processing pipeline tailored to commodity ANC-capable earphones.
Building on this, we repurpose classical LMS-based adaptive filtering to preserve in-ear respiration signals. We introduce \sysName{}, an integrated on-device pipeline for commodity earphones combining adaptive noise suppression, \RR{} estimation, and channel-discrepancy-based outlier handling.

%While some earphone-based systems exploit multiple microphones for speech enhancement, such as EarSpeech, these approaches rely on learned cross-channel representations optimized for speech intelligibility rather than on causal, reference-channel–based adaptive noise suppression. Unlike LMS- or FxLMS-based methods, EarSpeech does not perform physical noise cancellation but fuses channels through model-based learning~\cite{earspeech2024}.

\section{\sysName{}: System Overview}
\label{sec:approach}
In this section we introduce \sysName{}, a system for audio-based respiration rate estimation under environmental noise conditions. The system consists of three stages: (1) adaptive noise suppression (see \autoref{sec:ans}), (2) respiration rate estimation (see \autoref{sec:rr_est}), and (3) channel-discrepancy-based outlier rejection (see \autoref{sec:conf_est}). 

\begin{figure*}
    \centering
    \includegraphics[width=\linewidth,trim=0cm 7cm 0cm 0cm]{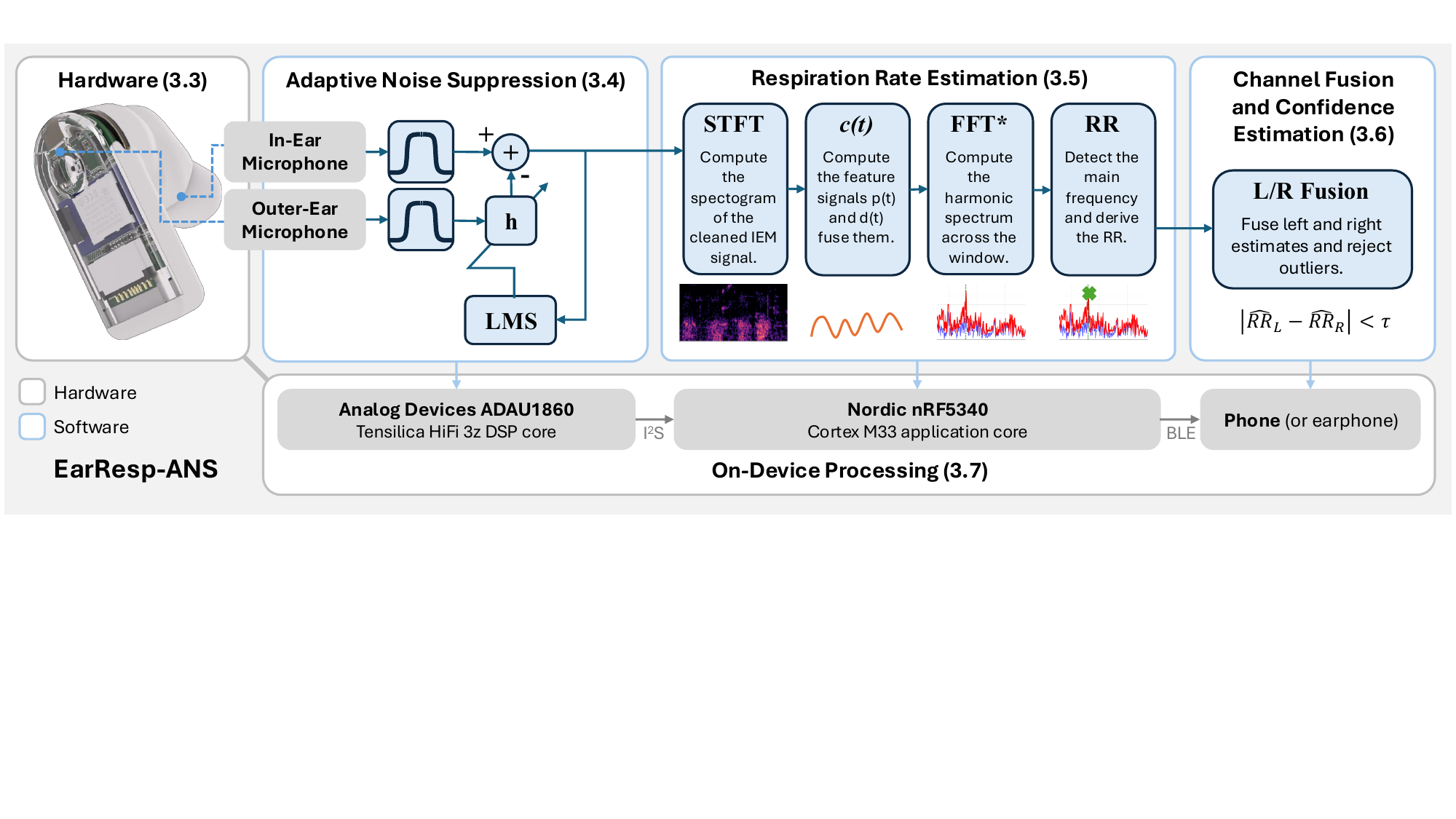}
    \caption{
    %System pipeline of \sysName{} on OpenEarable~2.0. ANS uses a delayed LMS filter on the DSP to denoise the IEM signal. \RR{} estimation (STFT-based features and peak selection) runs on the MCU; per-ear estimates are sent via BLE for fusion.
    System pipeline of \sysName{} on OpenEarable~2.0. ANS denoises the IEM signal on the DSP using a delayed LMS filter. \RR{} estimation (STFT-based features and peak selection) runs on the MCU, with per-ear estimates transmitted via BLE for fusion.}
    \label{fig:pipeline}
\end{figure*}

%\begin{figure}[t]
%\centering
%\input{figures/pipeline}
%\caption{System pipeline of \sysName{} deployed on the OpenEarable~2.0 platform. The figure illustrates the adaptive noise suppression (ANS) stage executed on the Digital Signal Processor (DSP), where a delayed LMS algorithm adapts the FIR filter~$h$. The noise-reduced signal is transmitted to the microcontroller unit (MCU) via I\textsuperscript{2}S, where respiration rate estimation is performed, including Short-Time Fourier Transform (STFT) computation, extraction of the feature signal $c(t)$, computation of the harmonic spectrum FFT*, and main frequency detection. Respiration rate estimates from each earable are transmitted to a smartphone via BLE, where channel fusion is applied.}
%\label{fig:pipeline}
%\end{figure}

%\todo[inline]{
%Design Goals, Hardware and Sensing Setup, Processing Pipeline, On-Device Execution and Privacy, System Output
%}

\subsection{Design Goals}
% energy saving 
The goal of this work is to design a respiration rate estimation system that can be executed entirely on-device on commodity ANC-capable earphones. In contrast to approaches that rely on streaming microphone signals to an external device, our method should be capable of operating solely on the earphone itself and should not require any raw audio data to leave the device.
To enable always-on, fully offline operation, the proposed algorithms are designed to meet the computational and memory constraints of commercial ANC-capable earphones, requiring only minimal processing and memory overhead.
% robustness to noise
In addition to these system-level constraints, the method should provide robust respiration rate estimates under realistic environmental noise conditions, such as background speech or background music playback. By avoiding microphone streaming over wireless interfaces, the proposed system should explicitly addresses privacy concerns associated with continuous audio sensing and simultaneously reduces energy consumption caused by radio transmission. 

\subsection{Adaptive Filtering Background}
\label{sec:background}
% This section is about adaptive noise suppression and gives an overview over appraoches, based on which we build the ANS appraoch later. we keep this section about IEM cleaning in general (not necessarily related to breathing sounds).

This section establishes the technical foundation for reference-signal-based approaches that employ adaptive filtering techniques suitable for earphone-based systems. Suppressing external noise in the in-ear microphone (IEM) signal is closely related to the problem of active noise cancellation (ANC), which aims to reduce the perceived noise level at the user’s eardrum by generating anti-noise. Several adaptive filtering techniques originally developed for ANC have therefore been successfully transferred to the task of numerically suppressing external noise in IEM signals, for example in the context of in-ear speech enhancement~\cite{bouserhal2017ear,KAJLA2020107035,SAYOUD2018101}.

% ANC sytems and need of adaptiveness
In classical ANC setups used in earphones, external noise is compensated by generating anti-noise through a loudspeaker. To effectively suppress external noise, both the acoustic path from the outer ear microphone (OEM) to the IEM, referred to as the primary path, and the path from the earphone loudspeaker to the IEM, referred to as the secondary path, must be accurately modeled.
However, the acoustic coupling between the earphones and the ear canal can vary substantially across users and may change over time due to differences in ear-tip fit or user motion. As a result, fixed or offline-tuned filters are often insufficient for accurate acoustic path estimation, motivating the use of adaptive approaches that continuously compensate for such variations~\cite{kuo2002active}.

%basic LMS
Adaptive filtering techniques such as the Filtered-x Least Mean Squares (FxLMS) algorithm~\cite{kuo2002active} are widely used in ANC systems. 
In its basic form, the LMS adaptive filter estimates the primary acoustic path from the OEM signal $x(n)$ to the IEM signal $d(n)$. The objective is to minimize the error signal $e(n)$ defined as
\begin{equation}
e(n) \coloneqq d(n) - \vect{h}^T \vect{x}(n)
\end{equation}
by adapting the filter $\vect{h}$ of length $M$, where $\vect{x}(n) = [x(n), x(n-1), \ldots, x(n-M+1)]^\top$ denotes the reference signal vector. This corresponds to minimizing the instantaneous squared error
\begin{equation}
J \coloneqq \left\lVert d(n) - \vect{h}^T \vect{x}(n) \right\rVert^2_2.
\end{equation}
% The following update rule
% \begin{equation}
% \frac{\partial J}{\partial \mathbf{h}}
% = -2\,\mathbf{x}(n)\,e(n).
% \end{equation}
Applying stochastic gradient descent yields the LMS weight update rule
\begin{equation}
\mathbf{h}(n+1) \coloneqq \vect{h}(n) + \mu\, e(n)\, \mathbf{x}(n),
\end{equation}
where $\mu$ denotes the adaptation step size.
% normalized LMS
To improve robustness under high-amplitude noise conditions, the gradient can be normalized by the instantaneous signal power of the OEM signal, resulting in the normalized LMS (NLMS) update rule
\begin{equation}
\vect{h}(n+1) \coloneqq \vect{h}(n) + \frac{\mu\, e(n)\, \vect{x}(n)}{\epsilon + \vect{x}^T(n)\vect{x}(n)},
\end{equation}
where $\epsilon$ is a small constant added for numerical stability. In addition, a leakage term $\gamma$ can be introduced to prevent weight drift and mitigate coefficient growth, leading to the leaky LMS formulation~\cite{tobias2005leaky}:
\begin{equation}
\vect{h}(n+1) \coloneqq (1 - \nu)\,\vect{h}(n) + \mu\, e(n)\,\vect{x}(n),
\qquad \nu = \gamma \mu.
\end{equation}

% transfer to adaptive noise reduction
Conventional ANC systems impose strict latency constraints, as the anti-noise signal must be generated with minimal delay to achieve destructive interference with the incoming noise. 
Instead of emitting anti-noise through a loudspeaker, adaptive filtering can also be applied to numerically suppress background noise directly in the IEM signal. In this setting, the adaptive filter estimates the noise component captured by the OEM and subtracts it from the IEM signal, thereby eliminating the need for secondary path modeling. Under the assumption that external background noise is uncorrelated with the signal of interest present in the IEM data, this approach enables effective isolation of the desired signal. Such reference-signal-based noise suppression techniques have been successfully applied, e.g., to IEM-based speech enhancement~\cite{bouserhal2017ear}.

% delayed LMS
When adaptive filtering is used for post-processing inner microphone signals for respiration rate estimation rather than for real-time sound cancellation, strict latency constraints are relaxed. Short delays on the order of several milliseconds are acceptable in this context. This enables the use of a delayed LMS adaptive filter for primary path estimation. By introducing a delay $K$ to the IEM signal, the strict causality requirement of the primary path is relaxed, allowing for a more stable and linear-phase filter response at the cost of increased group delay. The resulting error signal is defined as
\begin{equation}
e(n) \coloneqq d(n-K) - \vect{h}^T \vect{x}(n).
\end{equation}
In summary, these adaptive filtering formulations provide an efficient way to estimate and suppress the OEM-correlated noise component in the IEM signal. In the following, we build on this foundation and describe how we implement delayed LMS as an on-device adaptive noise suppression (ANS) stage tailored to earphone hardware.

\subsection{Hardware}\label{sec:hw_and_preprocess}
The proposed system is implemented on the OpenEarable~2.0 platform~\cite{roddiger2025openearable}. Audio signals are acquired using \emph{Knowles SPH0641LU4H-1} pulse-density modulation (PDM) microphones at a sampling rate of 48\,kHz from both the IEM and OEM. Both microphones are connected to an \emph{Analog Devices ADAU1860} audio DSP via a shared PDM bus.
The \emph{ADAU1860} features a dedicated \textit{FastDSP} core optimized for biquad filter operations, as well as a programmable \emph{Tensilica HiFi~3z} core based on the \emph{Xtensa} instruction set architecture~\cite{cadence_hifi3z}. The \emph{HiFi~3z} core supports up to four 24$\times$24 multiply--accumulate (MAC) operations per cycle\cite{cadence_hifi3z} and is operated at a clock frequency of 98.304\,MHz, enabling efficient real-time audio processing.
The DSP is connected to a \emph{Nordic nRF5340} system-on-chip, which features a 128\,MHz application core and a 64\,MHz network core.

\subsection{Adaptive Noise Suppression (ANS)} \label{sec:ans}
%intro sentence and motivation
As mentioned before, although the silicone ear-tips of earphones shield the IEM from outside noises to some extent, the recordings are dominated by environmental noise even for low noise levels.
To enable accurate respiration rate estimation, external noise components must be attenuated in the IEM signal. As discussed above, the acoustic transfer path from the OEM to the IEM varies across users, between insertions, and over time due to motion and changes in ear-tip fit. Therefore, \sysName{} estimates this transfer path adaptively.

We achieve this by employing a delayed LMS adaptive filter, as described in \autoref{sec:background}, to estimate the acoustic transfer path from the outer to the inner microphone. The filter is implemented as a \ntaps{}-tap FIR filter with a delay of \delay{} samples. The estimated noise component is subtracted from the inner microphone signal, yielding a noise-reduced respiration signal. 
%The filter coefficients are adapted to minimize the mean squared error between the inner microphone signal and the filtered outer microphone signal. As a result, the residual error signal, defined as the inner microphone signal minus the estimated noise contribution, contains minimal energy correlated with the external acoustic environment and predominantly preserves components that are uncorrelated with the external noise, such as respiration-related signals.

This approach relies on the assumption that respiration sounds are only weakly present in the outer microphone and are largely masked by environmental noise. Consequently, respiration-related components can be assumed to be uncorrelated with the reference signal and are therefore not removed by the adaptive filtering process. We discuss the validity and limitations of this assumption in \autoref{sec:discussion}.

To improve stability under high signal amplitudes,
we also apply a small leakage $\gamma$ to the filter weight as described in \autoref{sec:background}. We found that using normalized LMS reduces the performance as it lowers the gradient for louder noise.
Instead, we apply normalization only once the signal exceeds a certain threshold $\tau$, to prevent divergence on very large amplitudes. This yields the following update rule for the adaptive filter coefficients:
\begin{equation}
\begin{aligned}
\mathbf{h}(n+1) &\coloneqq (1 - \nu)\,\mathbf{h}(n) + \mu\,s_n\, e(n)\,\mathbf{x}(n), 
\qquad \nu = \gamma \mu, \\
e(n) &\coloneqq d(n-K) - \mathbf{h}^\top(n)\,\mathbf{x}(n), \\
\qquad s_n &\coloneqq \min\!\left(1,\dfrac{\tau}{\epsilon + e(n)\,\vect{x}^T(n)\vect{x}(n)}\right).
\end{aligned}
\end{equation}
where $\mathbf{x}(n)$ denotes the reference signal vector, $d(n)$ the inner microphone signal, $\mu$ the adaptation step size, and $\gamma$ the leakage factor. The error signal $e(n)$ denotes the noise suppressed inner microphone signal. We therefore define the noise reduced output signal as
\begin{equation}
    y(n) \coloneqq e(n).
\end{equation}

\begin{figure}[!t]
    \centering
    \includegraphics[width=\linewidth]{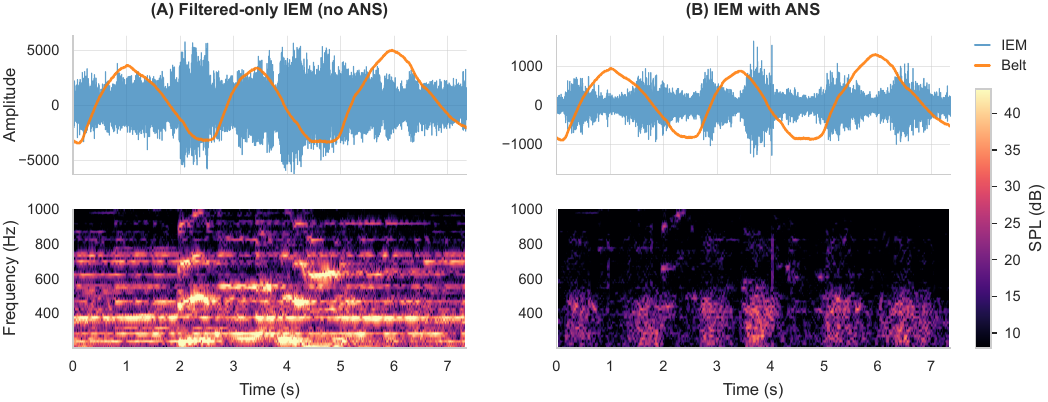}
    \caption{IEM signals and respiration belt ground-truth data before (A) and after (B) adaptive noise suppression, recorded during the music sound condition. The figure shows the raw time-domain signals together with their corresponding spectrograms. The inspiration and expiration segments vanish under the background noise in figure (A) and are recovered in figure (B) using ANS. Correlation with respiration belt signal becomes visible in both the raw audio signal and the spectrogram. }
    \label{fig:rr_sig}
\end{figure}

To focus on respiration-related acoustic components, both microphone signals are band-pass filtered between \fmin{} and \fmax{}, before applying the algorithm. This frequency range captures characteristic breath sounds while suppressing body noises and heartbeat sound on the low-frequency side, as well as higher-frequency environmental noise. We further investigated that the microphone data contains the highest correlation with the respiratory signal inside this range. 
It also allows for a lower sampling rate and thus yields a lower computational effort.
The effect of applying this algorithm to the microphone data can visually be observed in \autoref{fig:rr_sig}.

%The resulting algorithm is provided in Appendix (\autoref{alg:noise_suppression_vector}).

\subsection{Respiration Rate Estimation}  \label{sec:rr_est}

%After applying after applying the band-pass filter and the noise suppression, the signal is mainly dominated by breathing sound. Both the inhalation and exhalation will produce a 
We estimate the respiration rate over a fixed window size using two primary features: the logarithmic spectral signal energy and the spectral dissimilarity to a representative breathing pattern within the analysis window.

\subsubsection{Logarithmic Spectral Energy} \label{sec:spec_en}
We assume that after band-pass filtering and adaptive noise suppression, the inner microphone signal is dominated by respiration-related sounds. Consequently, the signal energy rises at each inspiration and expiration, since both generate prominent breath sounds, as illustrated in \autoref{fig:rr_sig}. This characteristic periodicity leads to a pronounced maximum in the spectrum of the energy signal.
We define the logarithmic spectral energy signal as
\begin{equation}
p(t) \coloneqq \frac{1}{M}\log\left(\sum_{f=0}^{F} \lvert Y(t,f)\rvert^2\right),
\end{equation}
where $Y(t,f)$ denotes the STFT time-frequency spectrogram with $F$ frequency bins of the cleaned IEM signal $y(t)$. The STFT is performed with a Hamming window of size \StftWinsize{} and a stride of \StftStride{}.

\subsubsection{Spectral Dissimilarity} \label{sec:spec_sim}
Respiration rate estimation is performed over a fixed-length analysis window. Under the assumption that the noise-suppressed and band-pass filtered signal is dominated by respiration-related acoustic components, we retain only time-frequency bins exceeding the 85\% quantile of spectral energy. This selective masking yields a robust estimate of the respiratory sound component by emphasizing high-energy regions that are most likely associated with breathing events.
The quantile threshold is chosen to balance two competing objectives: ensuring a high likelihood that the retained samples correspond to respiration-related activity, while preserving a sufficient number of samples to obtain a stable estimate of the average respiratory spectrum.
We empirically found this threshold to provide a good trade-off between robustness to noise and spectral stability across all evaluated conditions. Using the retained time--frequency bins, we estimate the average breath spectrum of each analysis window by normalizing individual STFT frames and computing their mean. For this normalization, we use the $p$-norm defined as
\begin{equation}
\norm{\vect{x}}_p
\coloneqq
\left(\sum_{n=0}^{N}{x_n^p}\right)^\frac{1}{p},
\end{equation}
for some vector $\vect{x}$.
Instead of using the maximum-norm (Tschebyschow-norm) for normalization, which can be defined by choosing $p=\infty$ with
\begin{equation}
\norm{Y(t, \cdot)}_\infty
\coloneqq \lim_{p \rightarrow \infty}{\norm{\vect{Y}(t, \cdot)}_p} =
\max_{f \in \{0,\ldots,F-1\}}{\abs{Y(t,f)}},
\end{equation}
we find that using $p=8$ provides the best trade-off between increased robustness to extreme spectral peaks and sufficient numerical stability. We thus obtain the average breath sound estimate as
\begin{equation} \label{eq:avg_breath}
\overline{\vect{Y}}
\coloneqq
\frac{1}{|\mathcal{Q}_{0.85}|}
\sum_{t\in\mathcal{Q}_{0.85}}
\frac{Y(t,f)}{\norm{\vect{Y}(t,\cdot)}_p},
%\qquad \mathcal{Q}_{q} \coloneqq \{t \in [0, N] | u \in [0, N] | p(u) < p(t) > \p(t) \}
\end{equation}
where $\mathcal{Q}_q$ denotes the set of time-frame indices whose spectral energy exceeds the empirical $q$-quantile within the current analysis window, i.e.,
\begin{equation}
\mathcal{Q}_q
\coloneqq
\left\{t \in \{0,\ldots,N-1\} \mid p(t) \geq \tilde{p}_q \right\},
\qquad
\tilde{p}_q
\coloneqq
\mathrm{quantile}_q\left(\{p(t)\}_{t=0}^{N-1}\right).
\end{equation}
For each time frame inside this window, the difference between the spectrum at the current frame and the average breath spectrum defined in \autoref{eq:avg_breath} is calculated. To reduce the influence of large spectral deviations and improve numerical stability, we apply a logarithmic compression to the resulting distance measure. The logarithm attenuates extreme peaks caused by transient spectral components while preserving relative differences for smaller deviations. 

Based on this, we define the spectral dissimilarity measure as
\begin{equation}
d(t)
\coloneqq
\frac{1}{F} \log\left(\sum_{f=0}^{F-1}{\left(\frac{Y(t,f)}{\norm{\vect{Y}(t, \cdot)}_p} - \overline{\vect{Y}}(f)\right)^2}\right)
\end{equation}

\subsubsection{Respiration Frequency Detection}
For the detection of the respiration frequency, the energy signal $p(t)$ and the spectral dissimilarity signal $d(t)$ are combined into a combined feature signal $c(t)$ given by
\begin{equation}
c(t)
\coloneqq
a_p \cdot \frac{p(t)}{\lVert \vect{p}(\cdot) \rVert_2}
-
a_d \cdot \frac{d(t)}{\lVert \vect{d}(\cdot) \rVert_2},
\qquad
\lVert \vect{x} \rVert_2
=
\sqrt{\sum_{k=0}^{N-1} x^2(k)},
\end{equation}
where we chose $a_p=a_d=0.5$ for equal weighting of both features. Notice that since $d(t)$ produces minima at each inspiration and expiration event, the sign of $a_d$ needs to be inverted to be properly fused with $p(t)$. 
To estimate the respiration rate from the resulting feature signal, we compute the Fourier transform of $c(t)$ using the fast Fourier transform (FFT) after applying a Hamming window. To get a better frequency resolution we perform a zero-padding in the time-domain before performing the FFT.

\begin{figure}[t]
    \centering
    \includegraphics[width=0.99\linewidth]{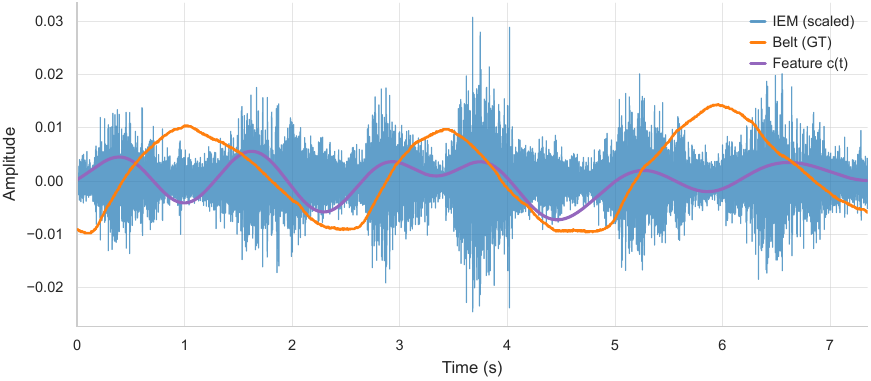}
    \caption{This figure shows the feature signal $c(t)$ compared to the denoised IEM signal. The IEM signal has been scaled for appropriate display. The feature signal contains local maxima for each inspiration and expiration. }
    \label{fig:feature_sig}
\end{figure}

% \todo[inline]{
%     Figure of diff signal compared to spectrogram \\
%     Figure of log spectral power signal compared to spectrogram \\
%     Figure of fused signal \\
% }

% \todo[inline]{
%     Analyze the spectrum of inhalation and exhalation. Show that for some participants the exhalation is more prominent then for others.
% }

Respiration produces two prominent spectral components, as both inspiration and expiration correspond to maxima in the feature signal. Consequently, spectral peaks appear at both the fundamental respiration frequency and its first harmonic (i.e., twice the breathing rate). The nature of the resulting signal can be observed in \autoref{fig:feature_sig}.
To enhance peak detectability, we therefore compute a harmonic spectrum defined as
\begin{equation}
C^{*}(f) \coloneqq \abs{C(f)} + \abs{C(2f)},
\quad f \in \{0,\ldots,\tfrac{F}{2}-1\}
\end{equation}
where $C(f)$ denotes the Fourier spectrum of the feature signal. The dominant peak in $C^*(f)$ is selected as the estimated respiration rate. We then obtain the estimated respiration rate as
\begin{equation}
    \RRhat{} \coloneqq \argmax_{f \in \{0,\ldots,\tfrac{F}{2}-1\}} C^{*}(f).
\end{equation}

\subsection{Channel Fusion and Outlier Rejection} \label{sec:conf_est}

Since our system consists of two earphones, each delivering a separate value for the estimated respiration rate, we can fuse both estimates to a single value to get a more accurate estimation. We fuse both channels by equal weighting, meaning that we obtain the fused value as
\begin{equation}
\label{eq:fusion}
\RRhat{}_{\mathrm{F}}
\coloneqq
\frac{\RRhat{}_{\mathrm{L}}}{2} + \frac{\RRhat{}_{\mathrm{R}}}{2}.
\end{equation}
Furthermore, we observe that windows exhibiting a larger absolute discrepancy between the left and right estimates, defined as
\begin{equation}
\label{eq:channel_dis}
\Delta\RRhat{} \coloneqq \lvert \RRhat{}_{\mathrm{L}} - \RRhat{}_{\mathrm{R}} \rvert,
\end{equation}
tend to yield higher estimation errors in the fused respiration rate $\RRhat{}_{\mathrm{F}}$. Based on this observation, we reject estimates for which $\Delta\RRhat{}$ exceeds a predefined threshold. The selection of this threshold is discussed in \autoref{sec:eval:conf_filter}, where we analyze the trade-off between estimation accuracy and the proportion of discarded windows.

%TODO: section

\subsection{On-Device Processing}

\sysName{} is implemented on the OpenEarable~2.0 platform and is partitioned across the on-board \tensilica{} DSP and the \nrf{} application processor. The ANS is executed on the DSP, which is optimized for low-latency audio preprocessing, while \RR{} estimation is performed on the MCU, which is designed for general-purpose on-device computation. This partitioning enables an efficient distribution of the computational workload between the two processing units. The complete processing pipeline is illustrated in \autoref{fig:pipeline}.

Microphone signals are acquired at a sampling rate of 48\,kHz and downsampled to 8\,kHz using the DSP’s decimator blocks. All subsequent noise suppression stages are executed on the \tensilica{} DSP core. Signal processing is performed in blocks of 24 samples using 24-bit fixed-point representation. As an initial preprocessing step, both the IEM and OEM signals are band-pass filtered using a fourth-order Butterworth filter, implemented as a cascade of biquad sections. The filter covers a passband from \fmin{} to \fmax{}, as described in \autoref{sec:hw_and_preprocess}.

Noise suppression is performed using a delayed LMS adaptive filter implemented on the Tensilica HiFi3z DSP, as described in \autoref{sec:ans}. The filter uses \ntaps{} FIR taps and applies a delay of \delay{} samples to the desired (inner microphone) signal. We again use 24-bit fixed-point representation, thus making use of the DSP's 24$\times$24 MAC units.

The processed audio stream is forwarded to the nRF5340 via an I\textsuperscript{2}S interface for subsequent feature extraction and respiration rate estimation. 
To reduce computational load while maintaining respiration-relevant information, the signal is further downsampled to a sampling rate of 2000\,Hz using a decimator chain. 
Respiration rate estimation is computed using an STFT-based pipeline implemented with the CMSIS-DSP library\cite{arm_cmsis_dsp} and the ARM DSP instruction set.

% more text
After the feature signal is computed and bandpass-filtered, we downsample it by factor 32 to approximately $3.9\,\mathrm{Hz}$, to remove unnecessary complexity and memory requirements for the harmonic spectrum computation. This still allows capturing respiration rates of up to $58.6\,\cpm{}$, which well covers the 99\%-quantile of breathing rates reported for sports activities such as running at a pace of 12\,km/h, thereby encompassing the vast majority of individuals~\cite{bios13060637}.

The resulting respiration rate estimate for each earable is then transmitted to a smartphone via Bluetooth Low Energy (BLE).
On the smartphone, respiration rate estimates are received from each earable and fused to obtain a single robust output as described in \autoref{eq:fusion}.
%Fusion can either be performed by equal weighting or based on the peak-mean-ratio as in \autoref{eq:peak-mean}, allowing the system to down-weight unreliable estimates under adverse acoustic conditions. \mk{maybe remove fusion}
Optionally, filtering of potential outliers can be performed based on the discrepancy between the estimates of the two channels, as described in \autoref{sec:eval:conf_filter}.
\section{Data Collection}
\label{sec:data}
This section describes the data collection procedure used to evaluate \sysName{}. We summarize the hardware platform and recording setup, the experimental protocol, and the derivation of ground-truth respiration rates. In addition, we describe the resulting dataset and the data cleaning and segmentation steps applied prior to analysis. The study protocol was designed to capture respiration signals under realistic acoustic conditions while ensuring reproducibility and reliable reference measurements.

\subsection{Experimental Protocol and Setup}

\begin{figure}[t]
    \centering
    \includegraphics[width=0.99\linewidth]{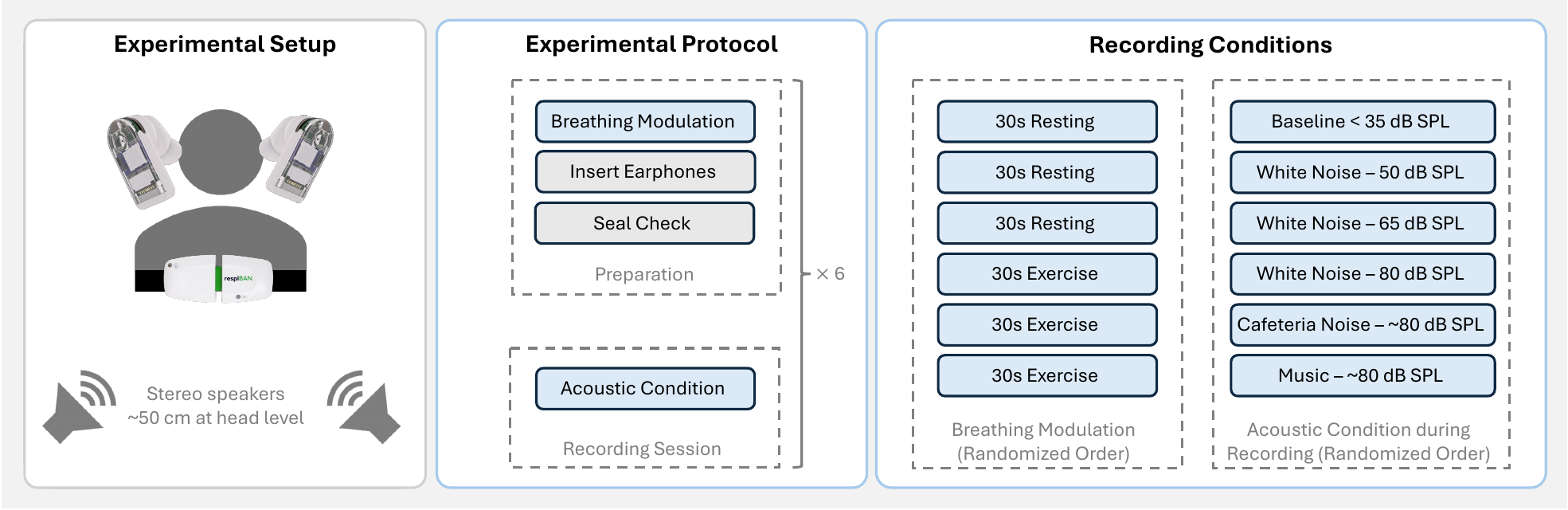}
    \caption{Experimental setup, recording protocol, and recording conditions. Participants wore binaural OpenEarable earphones and a respiBAN respiration belt while seated. Audio was played via stereo speakers (approx.\ 50\,cm, head level). Each participant completed six recording sessions in randomized order. For each session, breathing modulation (none or exercise-induced via 30\,s jumping jacks) and one acoustic condition were assigned. Acoustic conditions included a low-noise baseline (<35\,dB SPL), white noise (50/65/80\,dB SPL), cafeteria noise (approx.\ 80\,dB SPL), and music (approx.\ 80\,dB SPL); each condition was recorded once per participant. }
    \Description{Schematic of the seated study setup (earphones + respiration belt) and the session structure: six sessions per participant with randomized acoustic condition and optional exercise-induced breathing modulation.}
    \label{fig:feature_studyprotocol}
\end{figure}

% \todo[inline]{
% figure of setup
% }
\autoref{fig:feature_studyprotocol} provides an overview of the experimental setup and the recording protocol, including the sensor configuration, acoustic conditions, and session structure.
%Data were collected from a total of \numPart{} participants.
All participants wore OpenEarable smart earphones equipped with in-ear microphones for audio acquisition. In addition, a PLUX respiBAN BLE respiration belt~\cite{plux_respiban} was used as a reference system to obtain ground-truth respiration rate measurements by measuring the breath-related chest expansion. An example of the signal produced by this device is shown in \autoref{fig:feature_sig}. This system was chosen over alternate systems such as Continuous Positive Airway Pressure (CPAP) masks, as they influence the respiratory sound~\cite{VUONG2016281}.
For audio data collection the OpenEarable~2.0~\cite{roddiger2025openearable} platform was used, as described in \autoref{sec:hw_and_preprocess}. The microphone signals were downsampled to 8000\,Hz during recording using a cascaded decimation stage, with a second-order Butterworth anti-aliasing filter applied at each stage. The downsampled data were then recorded to a microSD card.

During all recordings, participants remained seated and were instructed to avoid bigger movements. However, to maintain a natural and comfortable posture over the recording duration, participants were allowed to read content on a smartphone or laptop, provided that no large movements were performed.

Prior to each recording, a seal check was performed to ensure proper occlusion of the ear canal. This was achieved by playing a multitone signal through the earphone speaker and analyzing the resulting frequency response measured by the in-ear microphone. The presence of pronounced low-frequency components was used to verify the occlusion, thereby ensuring consistent acoustic coupling between the device and the ear canal across recordings.
Recordings were conducted under the following acoustic conditions:
\begin{itemize}
    \item baseline condition with ambient noise levels below 35\,dB sound pressure level (SPL)
    \item white noise playback at 50\,dB, 65\,dB, and 80\,dB SPL
    \item cafeteria noise at approximately 80.0\,dB SPL ($\pm 5.2$\,dB SD over time), generated using artificial sound playback
    \item music condition at approximately 80.1\,dB SPL ($\pm 5.8$\,dB SD over time), consisting of pop music
\end{itemize}

All acoustic conditions were played back using commercially available stereo loudspeakers. The speakers were placed at head level at a distance of approximately 50\,cm from the participant to ensure consistent sound exposure across recordings. %\tr{I think a schematic drawing of the study protocol could still be helpful, however, it might be quite a small figure so idk if it looks good}

Each participant completed six recording sessions in total, corresponding to the six acoustic conditions. The order of the conditions was randomized across participants. To increase variability in the respiration rate, three randomly selected recording sessions per participant were recorded under exercise-induced breathing modulation, achieved via 30 seconds of jumping jacks immediately before the recording. Before each recording, the participants were instructed to remove and reinsert the earphones to increase variability in device fit within the ear canal.

All sound playback levels were calibrated prior to the study to ensure accurate and reproducible sound pressure levels across conditions. Calibration was performed using a sound level meter positioned at ear height at the participant location. The playback gain of the loudspeakers was adjusted such that the target sound pressure levels were achieved at a distance of 50\,cm from the sound source. 
For non-stationary acoustic conditions (cafeteria noise and music), calibration was performed using equivalent continuous sound pressure levels (LAeq). %During recordings, sound levels were periodically verified to ensure consistency and to compensate for potential deviations in playback.

\subsection{Ground Truth Derivation}

Evaluation is performed on \rrWin{} analysis windows. For each window, the ground-truth respiration rate is estimated by applying a FFT to the respiration signal using a Hamming window and extracting the dominant spectral peak. To improve frequency resolution, the signal is zero-padded to 32 times its original length prior to the FFT. This step is required because a window duration of \rrWin{} inherently yields a frequency resolution of $0.05$\,Hz (corresponding to $3$\,\cpm{}), which is insufficient for accurate respiration rate estimation. Zero-padding effectively interpolates the frequency spectrum, enabling a more precise localization of the dominant respiration frequency without altering the underlying signal content.

%A similar limitation arises when estimating respiration rate by counting peaks and valleys in the time-domain respiration signal. Since only an integer number of peaks and valleys can occur within a single window, this approach is inherently quantized and sensitive to window placement. In particular, whether a peak near the window boundary is included or excluded can significantly affect the resulting estimate, leading to increased variability across adjacent windows.

For data preprocessing, the continuous recordings were segmented into \rrWin{} windows with 50\% overlap. Windows with implausible respiration rates below 7.5\,\cpm{} or above 30\,\cpm{}, as determined from the ground-truth signal, were excluded from further analysis.
Additionally, windows for which no clear dominant peak corresponding to the respiration frequency could be identified in the FFT of the ground-truth respiration signal were removed. This step ensured that only segments with a reliable reference respiration rate were retained for evaluation. In total \winDropped{} of the frames were excluded due to unreliable ground truth signal.
%\tr{reviewers will most likely ask which fraction (percentage) of the dataset was excluded based on this process,}

\subsection{Dataset}

\begin{figure}[!t]
    \centering
    \includegraphics[width=0.99\linewidth]{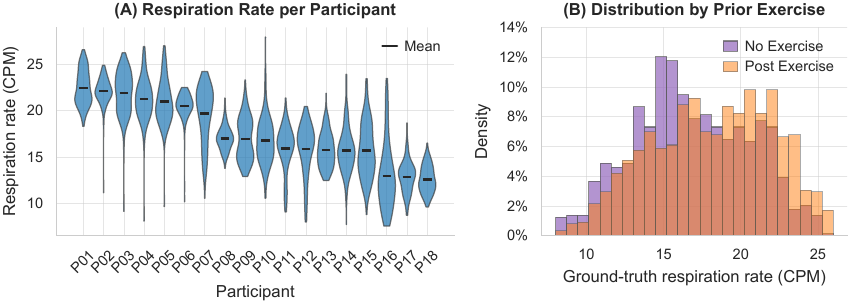}
    \caption{(A) Violin plot showing the distribution of ground-truth respiration rates for each participant. Participants are ordered post hoc according to their mean rate across all recordings. 
    %\tr{maybe you should explicitly state that they were ordered posthoc accoding to the mean breathing rate, otherwise this would be a crazy coincidence and a bit weird}. 
    (B) Histogram showing the distribution of ground-truth respiration rates with and without prior physical activity.}
    \label{fig:dataset_dist}
\end{figure}

The data were collected from 20 participants of which, two complete sessions were corrupted due to device failure of the ground truth signal, resulting in a total of \numWinsVal{} valid \rrWin{} windows from \numPart{} participants (Male: 14, Female: 6) of age between 22 and 46 years old ($\mathrm{M} = 26.55$, $\mathrm{SD} = 5.29$), with a breathing rate ranging between $7.57$\,\cpm{} and $27.94$\,\cpm{}. The distribution of breathing rates is illustrated in \autoref{fig:dataset_dist} for each participant as well as across activity conditions.
The resulting dataset shows a median respiration rate of $17.97$\,\cpm{} with a standard deviation of $3.84$\,\cpm{}. More detailed information can be found in the \autoref{sec:appendix} in \autoref{tab:activity}. Since fitness can affect how respiratory rate changes after activity, we documented participants’ usual exercise frequency: 9 reported no exercise, 5 exercised 1-2 times per week, 5 exercised 3-4 times per week, and 1 exercised more than four times per week.

% write about recording errors (left right)

\section{Evaluation} \label{sec:eval}

We evaluate \sysName{} with a focus on two aspects: (i) the overall performance of the respiration rate (RR) estimation algorithm, and (ii) the effectiveness of the noise suppression stage and its impact on respiration-related signal quality across different acoustic conditions. %An overview of the evaluation metrics and datasets used is provided in Section~\hl{TODO: ref eval section}.

\subsection{Respiration Rate Estimation Performance}\label{sec:rr_performance}
To evaluate the performance of the respiration rate estimation algorithm, we define the following metrics. All metrics are computed based on the estimated respiration rate, expressed in \cpm{}, and compared against the ground-truth reference.
A single respiration cycle is defined as one complete inspiration--expiration sequence. The duration of a cycle is measured from the onset of an inspiration to the onset of the subsequent inspiration. This definition is consistent with standard physiological interpretations of respiration rate~\cite{Liu_2019}.

As evaluation metrics we observe the Mean Absolute Error (MAE), defined as the absolute difference between the estimated respiration rate in \cpm{} and the ground truth, and Root Mean Square Error (RMSE), defined as the square root of mean squared error in CPM between the estimated respiration rate and the ground truth in \cpm{}.

\begin{table}[t]
\centering
\caption{Breathing-rate estimation error by condition.
Mean absolute error (MAE) and root mean square error (RMSE) are reported for all valid windows, as well as the estimation error when only using windows that are classified as confident (14.4\% dropped). For outlier detection, $\tau=0.52$\,\cpm{} has been chosen, which is the 3$\sigma$ interval limit of all inliers. 
For the single-channel and fused estimates, the corresponding inlier performance after
3$\sigma$ MAD outlier removal, as defined in \autoref{eq:i_mad}, is shown in parentheses.}
\label{tab:cond_metrics_combined_inline}
% \begin{tabular}{lcccccc}
% \toprule
%  & \multicolumn{3}{c}{\textbf{MAE (CPM)}} & \multicolumn{3}{c}{\textbf{RMSE (CPM)}} \\
% \cmidrule(lr){2-4} \cmidrule(lr){5-7}
% \textbf{Condition} 
%  & \textbf{Single (3$\sigma$)} & \textbf{Fused (3$\sigma$)} & \textbf{Confident}
%  & \textbf{Single (3$\sigma$)} & \textbf{Fused (3$\sigma$)} & \textbf{Confident} \\
% \midrule
% Baseline (<35\,dB)   
%  & 0.84 (0.13) & 0.75 (0.13) & 0.61
%  & 2.41 (0.17) & 2.14 (0.16) & 2.06 \\
% 50\,dB White Noise   
%  & 0.43 (0.11) & 0.41 (0.11) & 0.24
%  & 1.70 (0.14) & 1.50 (0.13) & 1.16 \\
% 65\,dB White Noise   
%  & 0.51 (0.12) & 0.50 (0.12) & 0.36
%  & 1.80 (0.16) & 1.63 (0.15) & 1.42 \\
% 80\,dB White Noise   
%  & 1.44 (0.15) & 1.33 (0.13) & 0.65
%  & 3.12 (0.20) & 2.71 (0.18) & 2.08 \\
% 65\,dB Cafeteria     
%  & 1.01 (0.16) & 0.93 (0.15) & 0.35
%  & 2.56 (0.21) & 2.11 (0.19) & 1.11 \\
% 80\,dB Music         
%  & 0.76 (0.15) & 0.74 (0.14) & 0.36
%  & 2.47 (0.19) & 2.29 (0.18) & 1.41 \\
% \midrule
% \textbf{Overall}     
%  & 0.82 (0.13) & 0.77 (0.12) & 0.42
%  & 2.38 (0.17) & 2.09 (0.16) & 1.57 \\
% \bottomrule
\begin{tabular}{lcccccc}
\toprule
 & \multicolumn{3}{c}{\textbf{MAE (CPM)}} & \multicolumn{3}{c}{\textbf{RMSE (CPM)}} \\
\cmidrule(lr){2-4} \cmidrule(lr){5-7}
\textbf{Condition}
 & \textbf{Single (3$\sigma$)} & \textbf{Fused (3$\sigma$)} & \textbf{Confident}
 & \textbf{Single (3$\sigma$)} & \textbf{Fused (3$\sigma$)} & \textbf{Confident} \\
\midrule
Baseline (<35\,dB)
 & 0.94 (0.15) & 0.85 (0.14) & 0.70
 & 2.52 (0.20) & 2.27 (0.19) & 2.21 \\
50\,dB White Noise
 & 0.52 (0.11) & 0.47 (0.11) & 0.29
 & 1.79 (0.14) & 1.56 (0.13) & 1.22 \\
65\,dB White Noise
 & 0.56 (0.14) & 0.55 (0.14) & 0.36
 & 1.85 (0.19) & 1.67 (0.18) & 1.35 \\
80\,dB White Noise
 & 1.54 (0.16) & 1.43 (0.15) & 0.75
 & 3.25 (0.22) & 2.84 (0.20) & 2.25 \\
65\,dB Cafeteria
 & 1.03 (0.18) & 0.95 (0.17) & 0.39
 & 2.50 (0.24) & 2.07 (0.22) & 1.18 \\
80\,dB Music
 & 0.85 (0.16) & 0.81 (0.15) & 0.41
 & 2.52 (0.20) & 2.30 (0.19) & 1.46 \\
\midrule
\textbf{Overall}
 & 0.90 (0.15) & 0.84 (0.14) & 0.47
 & 2.45 (0.19) & 2.16 (0.18) & 1.65 \\
\bottomrule
\end{tabular}
\label{tab:rr_performance}
\end{table}

\begin{figure}[t]
    \centering
    \includegraphics[width=0.99\linewidth]{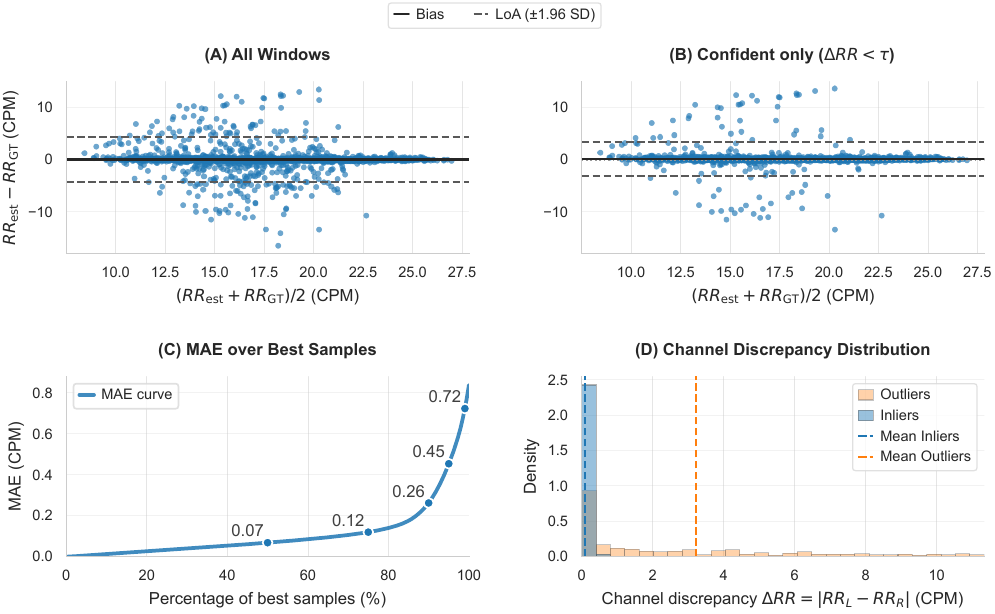}
    \caption{
    Analysis of fused respiration rate estimates and confidence-based filtering.
    \textbf{(A)} Bland--Altman plot (bias $\mu = 0.05\,\cpm$, $\mathrm{LoA}=[-4.18,\,4.27]\,\cpm$) for all valid analysis windows, showing the agreement between the fused estimate and the ground-truth respiration rate.
    \textbf{(B)} Bland--Altman plot (bias $\mu = 0.12\,\cpm$, $\mathrm{LoA}=[-3.11,\,3.34]\,\cpm$) after confidence filtering, retaining only windows with small inter-channel discrepancy ($\Delta RR < \tau = 0.52$\,\cpm{}).
    \textbf{(C)} Mean absolute error (MAE) as a function of the percentage of best samples, illustrating how estimation accuracy improves as increasingly unreliable windows are discarded.
    \textbf{(D)} Distribution of absolute inter-channel discrepancy $\Delta RR = |\mathrm{RR}_L - \mathrm{RR}_R|$ for inlier and outlier windows, highlighting the separation induced by the confidence criterion.}
    %\caption{Bland--Altman plot of \RR{} estimation using the fused output of \sysName{}. Red cross markers indicate estimates classified as low-confidence by the confidence estimation, while blue markers denote retained estimates.\tr{having the legend above the figure is weird, I suggest moving it below the figures}}
    \label{fig:bland-altman-rr}
\end{figure}

\subsubsection{Overall Performance}
% We first evaluate the performance of the respiration rate estimation algorithm using the noise-reduced in-ear microphone signals.
% All results reported in this section are obtained after applying the proposed adaptive noise suppression, which serves as a prerequisite for reliable respiration-related signal extraction.

% We analyze respiration rate estimation performance across all participants
% (\autoref{fig:bland-altman-participant}), comparing single-channel estimates, the fused estimation, and the effect of discarding low-confidence windows.
% An evaluation across different acoustic noise conditions is presented separately in \autoref{sec:ns_performance}, as it provides more direct insight into the effectiveness of the noise suppression stage rather than the respiration rate estimation itself.

We first evaluate the overall performance of \sysName{} for respiration rate (\RR{}) estimation using the noise-suppressed in-ear microphone signals. A detailed overview of the error metrics across all acoustic conditions is provided in \autoref{tab:rr_performance}, while the error distribution and agreement with the ground truth are visualized in \autoref{fig:bland-altman-rr} and \autoref{fig:kde-harmonics}.

Across all test conditions, the system achieves an MAE of \maeSingle{} for single-channel on-device estimation and \maeFuse{} after binaural fusion. As summarized in \autoref{tab:rr_performance}, applying channel discrepancy-based outlier detection (see \autoref{sec:conf_est}) further improves estimation accuracy, reducing the MAE to \maeConf{}.

The Bland--Altman plot in \autoref{fig:bland-altman-rr}~\textbf{(A)} corroborates these findings: the majority of estimates cluster closely around the zero-error line, indicating low systematic bias ($\mu = 0.05$) across the full observed breathing rate range from 7.57\,\cpm{} to 27.94\,\cpm{}. Channel discrepancy-based identification of unreliable windows effectively suppresses a large fraction of outliers, as illustrated in the post-filtering Bland--Altman plot shown in \autoref{fig:bland-altman-rr}~\textbf{(B)}. This results in a reduction of the limits of agreement from $\mathrm{LoA}=[-4.18,\,4.27]\,\cpm{}$ to $\mathrm{LoA}=[-3.11,\,3.34]\,\cpm{}$. A more detailed analysis of the outlier rejection strategy is provided in \autoref{sec:eval:conf_filter}.

\subsubsection{Performance across Participants}
\begin{figure}[t]
    \centering
    \includegraphics[width=0.99\linewidth]{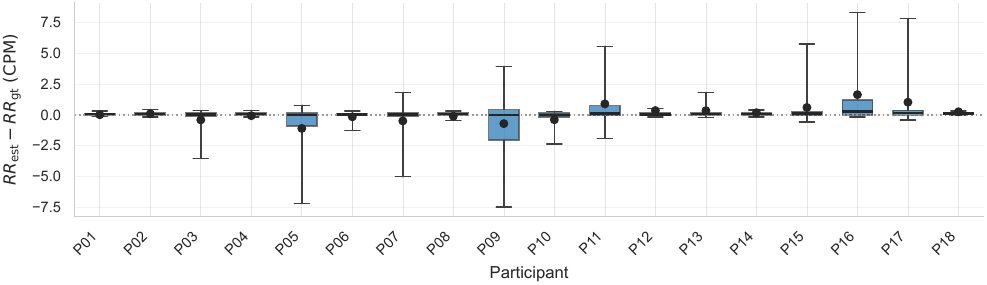}
    \caption{Distribution of \RR{} estimation error for \sysName{} across participants, shown as box plots.}
    \label{fig:box-participant}
\end{figure}

To assess generalization across participants, we analyze the signed prediction error using a variance decomposition approach that separates within-subject and between-subject variability.
Specifically, this decomposition quantifies the extent to which residual errors are driven by participant-specific characteristics.
Across all evaluation windows from the \numPart{} participants, the between-subject variance of the MAE amounts to $\sigma^2_{\text{between}} = 0.39\,\cpm{}^2$, while the within-subject variance is substantially larger at $\sigma^2_{\text{within}} = 3.96\,\cpm{}^2$.
This results in a generalizability ratio, defined as
\begin{equation}
    G \coloneqq \frac{\sigma_{\mathrm{between}}^2}{\sigma_{\mathrm{between}}^2 + \sigma_{\mathrm{within}}^2},
\end{equation}
of $G = 0.089$, indicating that less than $10\%$ of the total error variance can be attributed to subject-specific effects.
These findings demonstrate that, while inter-subject differences in mean error are present for individual participants, the dominant source of variability arises within participants.
Consequently, the proposed method exhibits strong generalization across users, with residual errors primarily driven by transient signal characteristics or recording conditions rather than systematic participant-dependent biases. The error distribution for each participant is visualized in \autoref{fig:box-participant}.

% \begin{figure}[t]
% \centering
% \begin{tikzpicture}[spy using outlines={circle, magnification=6, size=2cm, connect spies}]

% % Hauptgrafik
% \node[anchor=south west, inner sep=0] (img) 
%     at (-9.83,-0.4) {\includegraphics[width=0.99\linewidth]{figures/rr_bland_altman.pdf}};
% \begin{scope}[x={(img.south east)},y={(img.north west)}]

% % Zoom-Region definieren
% \spy[blue] on (0.45,0.55) in node[fill=white] at (0.85,0.25);

% \end{scope}
% \end{tikzpicture}
% \caption{Beispiel mit vergrößerter Detailansicht.}
% \end{figure}

\begin{comment}
\begin{figure}[t]
\centering
\begin{tikzpicture}

% --- Hauptbild ---
\node[anchor=south west, inner sep=0] (img) at (0,0)
    {\includegraphics[width=0.99\linewidth]{figures/rr_bland_altman.pdf}};

% Koordinatensystem auf Bild legen
\begin{scope}[x={(img.south east)}, y={(img.north west)}]

    % --- Kreis-Ausschnitt ---
    \begin{scope}
        \clip (0.65,0.55) circle (0.12); % Position + Radius
        \node at (0.65,0.55)
            {\includegraphics[width=0.25\linewidth]{figures/rr_kde_zoom_minus50.pdf}};
    \end{scope}

    % --- Kreis-Rand ---
    \draw[line width=0.8pt] (0.65,0.55) circle (0.12);

    % --- Verbindungslinie (optional) ---
    \draw[dashed] (0.65,0.55) -- (0.65,0.16);

\end{scope}
\end{tikzpicture}
\caption{Kreisförmige Vergrößerung eines Bildausschnitts.}
\end{figure}

\end{comment}

\subsubsection{Respiration Rate Estimation Performance Across Conditions}
\begin{figure}[t]
\centering
\includegraphics[width=0.99\linewidth]{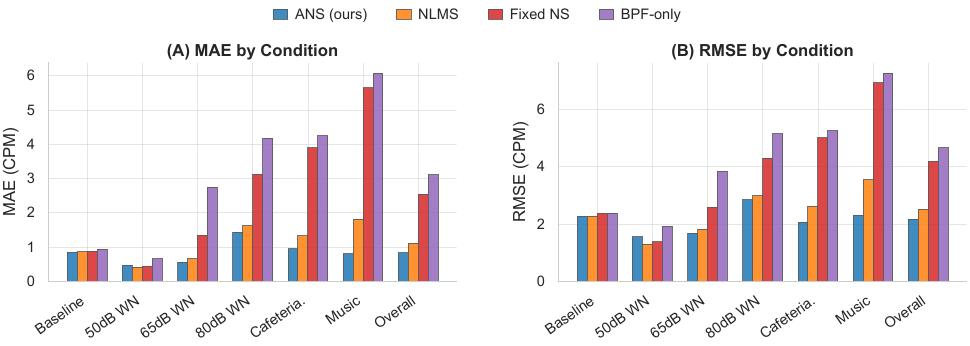}
\caption{Respiration rate estimation error using the estimation stage of \sysName{}, comparing different noise suppression approaches: ANS (ours), normalized LMS (NLMS), the offline-trained filter used in BreathPro~\cite{BreathPro2024} (Fixed NS), and a bandpass-filtered-only (BPF-only) baseline.}
\label{fig:cond_mae_rmse_bars}
\end{figure}

\autoref{fig:cond_mae_rmse_bars} summarizes respiration rate estimation errors across different acoustic conditions. More detailed results are provided in \autoref{tab:ns_rr_performance} in \autoref{sec:appendix}. We compare performance obtained from simple bandpass-filtered signals with results achieved using adaptive noise suppression (ANS). In addition, we benchmark our approach against the noise suppression method employed in BreathPro by \citet{hu2024breathpro}, which relies on a fixed filter trained offline across all participants. In BreathPro, noise reduction is performed in the frequency domain using an STFT-based filtering approach, after which the denoised signal is transformed back into the time domain for respiration rate estimation. In addition, we compare our delayed LMS approach against the normalized LMS method used by \citet{martin2002speechGammaDist}.
In order to evaluate the effectiveness of the various noise reduction methods in isolation, we used the same RR extraction pipeline from \sysName{} for all benchmarks, including the BreathPro approach. This ensures that the performance differences shown in \autoref{fig:mae_threshold} are due solely to the quality of the signal cleanup and not to variations in the estimation algorithm.

Overall, applying ANS results in a substantial performance improvement, reducing the mean absolute error (MAE) from $3.11$ to \maeFuse{}. For the baseline condition, both the noise-suppressed and filtered-only approaches achieve comparable performance (MAE ANS: $0.85$, MAE filtered: $0.94$).
In contrast, performance on the filtered-only data degrades significantly as noise levels increase, reaching an MAE of $6.07$ for the \Music{} condition. When ANS is applied, a substantial performance improvement is observed across all noisy conditions, with estimation accuracy remaining close to that of the baseline measurement (MAE $0.81$ for \Music{}). This demonstrates the robustness of the proposed approach to varying noise levels.
Performance is highest for \WNfifty{} (MAE $0.47$) and \WNsixty{} (MAE $0.55$), while slightly reduced performance is observed for music (MAE $0.81$) and cafeteria noise (MAE $0.95$). The lowest performance is obtained for \WNeighty{} (MAE $1.43$), consistent with the reduced signal quality (see \autoref{sec:ns_performance_corr}).%\mk{think about removing}

Comparing against the noise suppression approach of BreathPro~\cite{BreathPro2024}, our adaptive method consistently achieves better overall performance, yielding an MAE of \maeFuse{} compared to $2.54$. For the baseline condition, BreathPro performs comparably, but slightly worse than ANS (MAE ANS: $0.85$, MAE BreathPro: $0.88$). Under louder acoustic conditions, BreathPro performance degrades substantially. This effect is most pronounced for the music condition, where BreathPro reaches an MAE of $5.65$, compared to $0.81$ for the proposed adaptive approach.

The normalized LMS (NLMS) approach achieves performance comparable to that of the proposed ANS method, but consistently performs slightly worse across all conditions. In particular, for the music condition, the MAE is substantially higher (ANS MAE: $0.81$, nLMS MAE: $1.82$). This behavior can be attributed to the normalization, which improves numerical stability but simultaneously reduces the effective adaptation speed. As a result, the nLMS filter adapts more slowly to rapidly changing, non-stationary signals, leading to degraded performance in dynamic acoustic conditions.

\subsubsection{Impact of Physical Activity}
To evaluate the robustness of \sysName{} to changes in respiration dynamics, we analyze its performance in different physiological states: at rest and immediately following physical activity. Using the fused estimates, the algorithm achieves comparable accuracy across both conditions, demonstrating robust generalization over a wider range of \RR{} values.

Specifically, the fused MAE amounts to 0.61\,\cpm{} for measurements conducted after physical activity and 1.10\,\cpm{} during resting conditions. While the accuracy is slightly higher in the post-exercise state, this difference of 0.49\,\cpm{} remains well below the variability within each group, as reflected by the corresponding standard deviations (SD post exercise: 1.87\,\cpm{}, SD resting: 2.13\,\cpm{}). This indicates that the observed difference is not indicative of a systematic performance degradation at rest. The marginally improved accuracy following physical exertion may be attributed to more pronounced and acoustically salient breathing patterns, which can facilitate more reliable respiration-related signal extraction. Overall, these results confirm that the proposed method remains stable across different breathing rates. Further details are provided in \autoref{sec:appendix} in \autoref{tab:rr_activity}.

%To evaluate robustness to changes in respiration dynamics, we analyze performance before and after physical activity. Using the fused estimates, the algorithm achieves comparable accuracy across both conditions, demonstrating robust generalization over a wider range of respiration rates.

%Specifically, the fused MAE amounts to 0.54\,\cpm{} during sport activity\tr{misleading, it is never used during sport} and 0.84\,\cpm{} during no-sport conditions\tr{if find the naming misleading to call it no-sport condition and sport condition, they are both in the same setting so i doubt the activity itself has an influence, the sole reason to have}. While accuracy is slightly better during physical activity, this difference 0.30\,\cpm{} remains well below the variability within each group, as reflected by the corresponding standard deviations (sport: $\pm$1.64\,\cpm{}, no sport: $\pm$1.71\,\cpm{}). This indicates that the observed difference is not indicative of a systematic performance degradation at rest.

%The marginally improved accuracy after physical actibity may be attributed to more pronounced and acoustically salient breathing patterns following physical exertion, which can facilitate more reliable respiration-related signal extraction. Overall, these results confirm that the proposed method remains stable across different activity levels. Further details are provided in \autoref{sec:appendix} and \autoref{tab:rr_activity}.

\subsubsection{Distribution of Outliers}
\label{sec:eval:outlier}

Analyzing the distribution of the estimation error, we observe that the robust standard deviation estimate based on the median absolute deviation (MAD), $\hat{\sigma}=0.185$, is substantially smaller than the conventional standard deviation of \maeFuseSD{}. This indicates a heavy-tailed error distribution dominated by a small number of outliers, which can be observed in \autoref{fig:bland-altman-rr}~\textbf{(C)}.

Overall, 80.3\% of all estimates fall within a robust three-sigma equivalent interval centered around the median, corresponding to an absolute error below $0.612$. Restricting the evaluation to this confidence interval, defined as
\begin{equation}
\label{eq:i_mad}
\mathcal{I}_{\mathrm{MAD}}
\coloneqq
\left[
\mathrm{median}(x)
-
3 \hat{\sigma},
\;
\mathrm{median}(x)
+
3 \hat{\sigma}
\right],
\quad
\hat{\sigma}
\coloneqq
1.4826 \cdot
\mathrm{median}\!\left(
\left|x_i - \mathrm{median}(x)\right|
\right),
\end{equation}
reduces the MAE from \maeFuse{} to $0.14$.

\begin{figure}[t]
    \centering
    \includegraphics[width=0.99\linewidth]{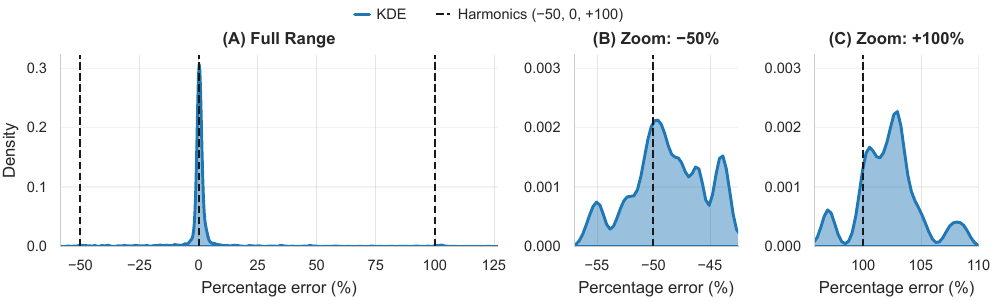}
    \caption{
    Distribution of percentage respiration rate estimation error.
    \textbf{(A)} shows the kernel density estimate (KDE) of the percentage error over the full range of observed values.
    The two right panels provide magnified views of prominent harmonic structures centered around \(-50\%\) \textbf{(B)} and \(+100\%\) \textbf{(C)}, respectively.
    %All KDEs are computed using identical bandwidth parameters, and the zoomed panels share a common density scale to enable direct visual comparison.
    }
    \label{fig:kde-harmonics}
\end{figure}

Further inspection of the outlier distribution reveals characteristic clustering around twice and half of the true respiration rate. This effect is reflected in the percentage error distribution shown in \autoref{fig:kde-harmonics} as distinct clusters near $-50\%$ and $100\%$. These clusters indicate harmonic estimation errors, where the algorithm erroneously selects a subharmonic or harmonic component instead of the fundamental respiration frequency.

\subsubsection{Channel Discrepancy-Based Outlier Rejection}\label{sec:eval:conf_filter}
\begin{figure}[t]
    \centering
    \includegraphics[width=0.99\linewidth]{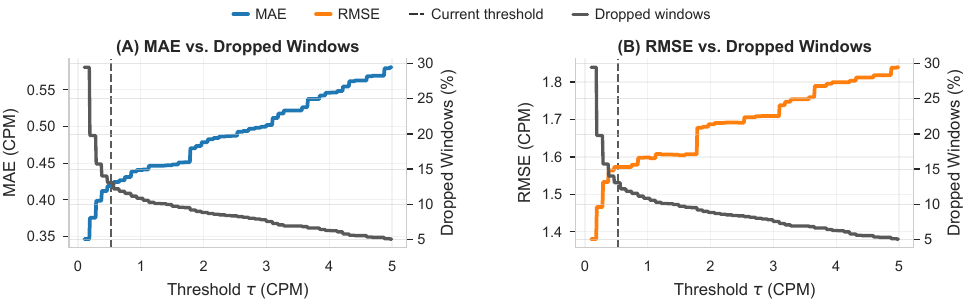}
    \caption{The graph shows the relationship between the threshold $\tau$ used for channel-discrepancy-based outlier removal, defined as $\Delta\RRhat{} = \lvert \RRhat{}_{\mathrm{L}} - \RRhat{}_{\mathrm{R}} \rvert$ (\autoref{sec:conf_est}), and its impact on the resulting MAE \textbf{(A)} and RMSE \textbf{(B)}.}
    \label{fig:mae_threshold}
\end{figure}

Finally, we analyze the relationship between channel discrepancy $\Delta\RRhat{}$, as defined in \autoref{eq:channel_dis}, and the occurrence of outliers. Specifically, we compute the absolute difference between the respiration rate estimates obtained from the left and right earphones. We observe that analysis windows with larger inter-channel discrepancies tend to exhibit higher estimation errors, as illustrated in \autoref{fig:bland-altman-rr} (D).
This finding supports the use of channel agreement as a confidence indicator and motivates its application for outlier suppression. However, some outliers still exhibit only small channel discrepancies, indicating that this criterion alone cannot eliminate all erroneous estimates. This behavior is also visible in the Bland--Altman plots before and after outlier rejection shown in \autoref{fig:bland-altman-rr}, where the majority of outliers are suppressed but a small number remains.
Despite this limitation, the proposed outlier rejection strategy yields a substantial improvement in estimation accuracy, reducing the MAE from \maeFuse{} to \maeConf{}. The influence of the discrepancy threshold on both the estimation error and the proportion of rejected windows is further illustrated in \autoref{fig:mae_threshold}.

\subsection{Noise Suppression Performance} \label{sec:ns_performance}

We first investigate the benefit of applying the delayed LMS-based noise reduction to the inner microphone signal. Noise suppression performance is evaluated both in terms of signal energy reduction and the preservation of respiration-related information. For measurement of preservation of respiratory information, we monitor the correlation with the respiratory signal as well as the performance of the approach on the data after noise reduction.

\subsubsection{Noise Reduction}

To quantify noise reduction, we compare the signal energy of the inner microphone recordings before and after applying the delayed LMS filter. Signal energy is computed over the respiration-relevant frequency band for each analysis window. The noise reduction is thereby given as:
\begin{equation}
\mathrm{NR}
=
10 \log_{10}
\left(
\frac{\sum_{n=0}^{N-1} e^2(n)}{\sum_{n=0}^{N-1} d^2(n)}
\right)
\;\mathrm{dB}.
\end{equation}

% \begin{figure}[t]
% \centering
% \begin{subfigure}[t]{0.49\linewidth}
% \centering
% \begin{tikzpicture}
% \begin{axis}[
%     ybar,
%     bar width=4pt,
%     width=\linewidth,
%     height=4.2cm,
%     ymin=0,
%     ymax=0.55,
%     ylabel={Correlation (median)},
%     symbolic x coords={
%         Baseline,50dB WN,65dB WN,80dB WN,65dB Caf.,80dB Music
%     },
%     xtick=data,
%     x tick label style={rotate=35,anchor=east,font=\scriptsize},
%     y tick label style={font=\scriptsize},
%     ylabel style={font=\scriptsize},
%     legend style={
%         font=\scriptsize,
%         at={(0.5,1.18)},
%         anchor=south,
%         legend columns=2,
%         /tikz/every even column/.append style={column sep=6pt},
%         draw=none
%     },
%     grid=both,
%     grid style={line width=.1pt, draw=gray!30},
%     major grid style={line width=.2pt, draw=gray!40},
%     enlarge x limits=0.10,
% ]
% % Filtered
% \addplot coordinates {
%     (Baseline,0.3920) (50dB WN,0.3626) (65dB WN,0.0702)
%     (80dB WN,0.0137) (65dB Caf.,0.0200) (80dB Music,0.0006)
% };
% % NS (ANS)
% \addplot coordinates {
%     (Baseline,0.3909) (50dB WN,0.4698) (65dB WN,0.4688)
%     (80dB WN,0.3862) (65dB Caf.,0.4441) (80dB Music,0.4346)
% };
% \legend{Filtered,ANS}
% \end{axis}
% \end{tikzpicture}
% \end{subfigure}

% \caption{Correlation comparison across acoustic conditions.}
% \label{fig:cond_corr_bars}
% \end{figure}

\begin{table}[t]
\centering
\caption{Median-based correlation and noise-reduction analysis across acoustic conditions.
Correlation metrics are reported for the filtered signal, the ANS signal, and their ratio (NS / Filtered).
Noise reduction is reported as the energy reduction $1-(\mathrm{E_{ns}}/\mathrm{E_{filter}})$ and the
corresponding attenuation in dB computed as $10\log_{10}(\mathrm{E_{ns}}/\mathrm{E_{filter}})$.}
\label{tab:ans-results}
\begin{tabular}{lccccc}
\toprule
 & \multicolumn{3}{c}{\textbf{Correlation (median)}} &
 \multicolumn{2}{c}{\textbf{Noise Reduction}} \\
\cmidrule(lr){2-4} \cmidrule(lr){5-6}
\textbf{Condition} &
\textbf{Filtered} &
\textbf{ANS} &
\textbf{Ratio} &
\textbf{[\%]} &
\textbf{[dB]} \\
\midrule
Baseline (<35\,dB)
 & 0.31 & 0.31 & 1.0
 & 0.6 & $-0.03$ \\

50\,dB White Noise
 & 0.33 & 0.41 & 1.2
 & 23.4 & $-1.2$ \\

65\,dB Cafeteria
 & 0.01 & 0.38 & 30
 & 98.9 & $-19.6$ \\

65\,dB White Noise
 & 0.06 & 0.44 & 7.4
 & 86.7 & $-8.8$ \\

80\,dB White Noise
 & 0.01 & 0.36 & 33
 & 99.1 & $-20.4$ \\

80\,dB Music
 & 0.01 & 0.38 & 44
 & 98.6 & $-18.5$ \\
\bottomrule
\end{tabular}
\end{table}

\subsubsection{Respiration-Related Signal Correlation} 
\label{sec:ns_performance_corr}

In addition to noise reduction, it is essential that respiration-related information is preserved. Physiologically, the speed of chest movement is correlated with airflow velocity, which in turn modulates the amplitude of respiration sounds captured by the inner microphone. Consequently, the temporal envelope of the respiration-related signal is expected to correlate with the slope of the ground-truth respiration signal.
To capture this relationship, we compute the correlation between the mean energy of the microphone signal and the absolute slope of the ground-truth respiration signal, where we compute the secant slope using $K=100$, which translates to a $0.25\,\mathrm{s}$ time-delay at $400\,\mathrm{Hz}$ sampling rate. Based on this, we formally define a respiratory information (RI) index as an evaluation metric as

\begin{equation}
\begin{aligned}
\mathrm{RI} &\coloneqq 
\mathrm{corr}\!\left(E_y, s_g\right) \\
E_y(t) &\coloneqq \frac{1}{T} \sum_{k=0}^{K-1} y^2(t-k) \\
s_g(t) &\coloneqq \frac{\abs{g(t) - g(t-K)}}{N},
\end{aligned}
\end{equation}

\noindent where $corr(x,y)$ denotes the correlation between $x$ and $y$.

\subsubsection{Results}
The quantitative results are summarized in \autoref{tab:ans-results}.
% reduction
Across all acoustic conditions, we observe a consistent reduction in signal energy after applying noise suppression, indicating effective attenuation of external noise components. The measured energy reduction ranges from $-0.03\,\mathrm{dB}$ for the baseline condition (\Base{}) to $-20.4\,\mathrm{dB}$ for \WNeighty{}.

% filtered only
Before applying noise suppression, a notable correlation between the inner microphone signal and the respiration signal is only observed for the baseline condition ($0.31$) and the \WNfifty{} measurement ($0.33$). For the remaining conditions with higher sound pressure levels, only weak correlation with the respiration signal is present. This effect is particularly pronounced for the music condition, where the RI drops to $0.01$, most likely due to the strongly non-stationary nature of the noise.
% correlation
After applying ANS, we observe consistently higher correlation values above 0.31 across all acoustic conditions. This demonstrates that the adaptive filter effectively preserves respiration-related signal components. Notably, the resulting correlation values remain comparable to the baseline measurement ($0.31$), indicating that the preserved signal quality after noise removal is on par with near-ideal recording conditions.
% RI
For all acoustic conditions except the baseline, noise suppression substantially increases respiratory information, ranging from factor $1.2$ (\WNfifty{}) up to factor $44$ (\Music{}). These results highlight the effectiveness of the proposed approach in enhancing respiration-related information under adverse acoustic conditions.
The highest signal quality is achieved under stationary acoustic conditions, particularly for white noise at 50\,dB and 65\,dB. In contrast, the lowest signal quality is observed for \WNeighty{}. In this condition, the noise-reduced signal retains less than $1\%$ of the original signal energy, indicating that the respiration signal is strongly dominated by external noise. This effect likely contributes to the slightly reduced respiration rate estimation performance observed under this condition compared to the other white noise scenarios.

Taken together, these results indicate that fixed, offline-trained filters do not generalize sufficiently acoustic conditions across different ear-tip fits. While such filters may perform adequately under quiet or mildly noisy scenarios, they fail to robustly recover respiration-related signals in the presence of stronger and non-stationary noise. In contrast, the proposed adaptive noise suppression (ANS) consistently preserves respiration-related information across all evaluated conditions, highlighting the importance of online adaptation for robust and participant-independent in-ear respiration sensing.

\subsection{Ablation Study}

The contribution of each system component to the overall performance is summarized in \autoref{tab:ablation}. The results corroborate the observations reported in the previous sections. Under the recording conditions considered in this study, ANS has the largest impact on performance, reducing the MAE from $3.26$ to $\maeSingle{}$. 
Furthermore, combining the log-spectral energy feature $p(t)$ with the spectral dissimilarity feature $d(t)$ consistently outperforms configurations that rely on either feature alone, indicating that the two representations capture complementary respiration-related information. 
Additional performance gains are achieved by fusing the respiration rate estimates from both earphones. However, the most substantial improvement beyond ANS is obtained by discarding low-confidence windows, which further reduces the MAE from $\maeFuse{}$ to $\maeConf{}$.

\begin{table}[t]
\centering
\caption{Ablation study analyzing the impact of ANS, feature composition $c(t)$, channel fusion, and confidence filtering on respiration rate estimation performance.}
\label{tab:ablation}
\begin{tabular}{lccccc c}
\toprule
\textbf{ANS} & \multicolumn{2}{c}{\textbf{Feature $c(t)$}} & \textbf{Fusion} & \textbf{Conf.} & \textbf{MAE (CPM)} & \textbf{RMSE (CPM)} \\
\cmidrule(lr){2-3}
            & $\mathbf{d(t)}$ & $\mathbf{p(t)}$            &                 &               &                    &                      \\
\midrule
$\times$     & \checkmark      & \checkmark                 & $\times$        & $\times$      & 3.26               & 5.01                 \\
\checkmark   & \checkmark      & $\times$                   & $\times$        & $\times$      & 1.10               & 2.79                 \\
\checkmark   & $\times$        & \checkmark                 & $\times$        & $\times$      & 0.90               & 2.55                 \\
\checkmark   & \checkmark      & \checkmark                 & $\times$        & $\times$      & \maeSingle{}        & \rmseSingle{}         \\
\checkmark   & \checkmark      & \checkmark                 & \checkmark      & $\times$      & \maeFuse{}          & \rmseFuse{}           \\
\checkmark   & \checkmark      & \checkmark                 & \checkmark      & \checkmark    & \textbf{\maeConf{}} & \textbf{\rmseConf{}}  \\
\bottomrule
\end{tabular}
\end{table}

\subsection{Runtime Performance and Latency}

\begin{table}[t]
\centering
\caption{Memory footprint, real-time index (RT index), and latency (processing induced and through buffer delay) of the individual system components, which are the ANS performed on the DSP, the computation of the STFT and the feature signal extraction together with the \RR{} estimation performed at a rate of 1\,Hz.}
\label{tab:system_resources}
\begin{tabular}{lcccc}
\toprule
\textbf{Component} &
\textbf{Memory} &
\textbf{RT Index} &
\textbf{Comp. Latency} &
\textbf{Buffer Latency} \\
\midrule
ANS            & 3488\,Byte  & 1.20\,\% & -- & 11\,ms \\
STFT           & 768\,Byte   & 1.56\,\% & 122\,$\mu$s &  8\,ms \\
Feature \& RR  & 12000\,Byte & 0.10\,\% & 1007\,$\mu$s & 1000\,ms \\
\bottomrule
\end{tabular}
\label{tab:rt-performance}
\end{table}

We evaluate the runtime performance and end-to-end latency of the proposed system for both the ANS stage executed on the DSP and the respiration rate estimation pipeline executed on the MCU. The results are summarized in \autoref{tab:rt-performance}.

\subsubsection{DSP Runtime and Audio Latency}

The ANS algorithm is executed on the \tensilica{} DSP at a sampling rate of 8000\,Hz, corresponding to a four-times oversampling relative to the respiration-relevant frequency band. Under this configuration, the DSP load introduced by the delayed LMS filter remains low, accounting for approximately 1.2\% of the available processing capacity.
Audio processing is performed in blocks of 24 samples, resulting in an intrinsic processing latency of $24$ samples at 8000\,Hz, corresponding to 3\,ms. In addition, the delayed LMS filter introduces an explicit delay of 64 samples (8\,ms) to ensure causality of the estimated primary path. The additional processing time is negligible compared to the block-based buffering delay. Together, this results in a total audio latency of approximately 11\,ms for the noise-reduced microphone signal.
Including the filter coefficients as well as the required input and gradient buffers, the ANS stage requires approximately 3.5\,kB of memory.

\subsubsection{MCU Runtime for Respiration Rate Estimation}

Respiration rate estimation is executed on the \nrf{} MCU using a downsampled microphone signal at 2000\,Hz. The short-time Fourier transform (STFT) is computed using a window length of 128 samples, a hop size of 16 samples, and 16-bit fixed-point precision. Under this configuration, the STFT computation exhibits a mean real-time factor of 1.56\%, corresponding to an average execution time of approximately 122\,$\mu$s per STFT frame. Including all input and output buffers required for FFT computation and windowing, the memory footprint of this stage amounts to 768\,bytes.
The window-wise computation of the harmonic spectrum using 32-bit floating point precision, including peak detection and respiration rate estimation, is performed at a rate of 1\,Hz. This stage incurs an additional computational load of 0.10\%, corresponding to an average execution time of approximately 1007\,$\mu$s per update. Storing the downsampled STFT results and the corresponding feature signals over a \rrWin{} window and extracting the feature signals requires a total of 12\,kB of memory.

\subsubsection{Real-Time Suitability}

Overall, the measured runtime and latency demonstrate that the proposed system operates well within real-time constraints on commodity embedded hardware.
The algorithm introduces only negligible computational overhead on both the MCU and the DSP, corresponding to an average utilization of 1.56\% and 1.20\% of the available processor time budget, respectively.
The approach further requires only $2.50\%$ of the MCU RAM (512\,KB)~\cite{nordic_nrf5340} and $1.56\%$ of the DSP's RAM (224\,KB)~\cite{analog_adau1860} .
These results confirm the feasibility of fully on-device, real-time respiration monitoring without compromising system responsiveness or resource availability for additional sensing tasks.

%\todo{Adaptation speed of ANS?}
\section{Discussion}
\label{sec:discussion}

This work establishes a foundation for unobtrusive, fully on-device respiration sensing on commodity earphones, enabling both technical extensions of the sensing pipeline and application-driven exploration of relaxation- and health-oriented use cases. 
\sysName{} is explicitly designed for respiration monitoring in low-motion contexts such as rest, recovery, or relaxation scenarios.
While \sysName{}
%the proposed system 
demonstrates robust performance across a wide range of acoustic conditions, several limitations must be considered. In the following, we discuss these limitations and outline potential mitigation strategies as well as promising directions for future work.

\textbf{Performance under Motion.} A primary limitation of the current evaluation concerns motion.
Data collection was conducted in a seated position with only minimal movement allowed. As a result, the presented evaluation does not fully capture the impact of stronger motion artifacts, such as those caused by walking or running. While the proposed system is well suited for monitoring respiration during rest or recovery phases (e.g., post-exercise cooldown), its performance during active movement remains an open question and requires further investigation.

\textbf{Impact of Seal on Occlusion-Effect.}
Beyond motion, signal quality is strongly influenced by the physical coupling between the earphone and the ear canal. 
During data acquisition, a proper seal of the ear canal was verified after participants inserted the earphones. In real-world use, however, an imperfect seal may reduce both the occlusion effect and the physical isolation of the IEM from external noise sources. Although average external noise levels in everyday scenarios are typically lower than in extreme conditions exceeding 80\,dB SPL, it remains unclear whether respiration-related acoustic components could become too weak relative to residual background noise, potentially approaching the sensitivity limits of the microphones and thereby reducing the reliability of signal recovery using the noise suppression algorithm.
However, it is worth emphasizing that the seal check in this work was performed automatically by measuring the earphone speaker’s in-ear frequency response via the IEM. This suggests that improper fit could be detected by the system and potentially accounted for in practical deployments.

\textbf{Assumption of Dominance of Respiration-Related Signals.}
The proposed approach assumes that respiration-related acoustic components dominate the noise reduced IEM signal after applying ANS. When this assumption is violated---for example due to an improper ear canal seal or high-amplitude disturbances---residual noise may dominate the signal, leading to unreliable respiration rate estimates.
This limitation could be addressed by monitoring the spectral consistency of the extracted breath-related segments within each analysis window. By comparing these segments to an average respiratory spectrum template, windows dominated by residual noise can be suppressed. Alternatively, the proposed method could be fused with complementary respiration rate estimation techniques based on other sensing modalities.
This assumption defines the operating regime of the proposed system: reliable respiration estimates are only possible when respiration-related components are sufficiently dominant in the filtered signal. When this condition is violated, estimation errors and outliers naturally increase, as observed in our evaluation and discussed below.

\textbf{Outlier Rejection.}
Independently of the specific failure modes discussed above, the observed error distribution in \autoref{sec:eval:outlier} indicates that a small fraction of outlier windows has a disproportionate impact on overall performance. These outliers may arise from a combination of residual noise, fitting-related effects, and inaccuracies in the reference signal.
%Building on the evaluation in \autoref{sec:rr_performance}, the observed error distribution shows that removing a small fraction of outlier windows substantially improves overall performance, reducing the MAE even more. 
While the proposed confidence estimation based on channel discrepancy already filters a subset of unreliable windows, this mechanism cannot eliminate all erroneous estimates. This motivates future work on extended confidence-aware and multimodal fusion strategies, for example by combining the proposed audio-based approach with complementary modalities such as IMU- or PPG-based respiration sensing, to further improve robustness under transient disturbances and challenging acoustic conditions.

\textbf{Validity of the Ground Truth.}
In addition to system- and algorithm-level limitations, the validity of the reference signal must be considered when interpreting the reported results. 
A total of \winDropped{} analysis windows were discarded due to unreliable ground-truth respiration estimates, identified by the absence of a clear breathing pattern. Potential causes include variations in belt tension around the chest or transient disturbances such as accidental contact with the sensor by participants, which can interfere with the measurement and compromise the reliability of the reference signal.
Our analysis indicates that further filtering of ground-truth windows based on the prominence of the respiration frequency leads to a substantial reduction in outliers and an apparent increase in overall estimation performance. This observation suggests that a portion of the remaining outliers might originate from inaccuracies in the ground-truth extraction rather than failures of the proposed respiration rate estimation algorithm itself.

\textbf{Performance under Baseline Condition.} We observe a slight decrease in correlation between the IEM signal amplitude and the ground-truth respiration signal when noise suppression is applied to some of the baseline recordings. One possible explanation is that under very quiet conditions, respiration-related components may also be present in the outer microphone signal and are therefore partially removed by the adaptive filter. Alternatively, the adaptive filter may not sufficiently adapt when little external noise is present, leading to reduced effectiveness for short, transient disturbances. These effects highlight that adaptive noise suppression is most beneficial in the presence of external noise and may be less advantageous under near-silent conditions. Future work could evaluate a level-dependent control or a bypass mechanism for the ANS module in order to optimize signal integrity in near-silent environments by proactively avoiding over-filtering of the breath components present in the external microphone.
%Furthermore, we notice that the performance is worse compared to the other conditions and performing worse than on the \WNfifty{} unsuppressed data, which indicates the quality of the recordings might be worse, or the ground-truth data may be partially corrupted. \mk{should we be this honest?}

\textbf{Music-Playback via Earphones.} 
Beyond the evaluated sensing scenarios, additional system extensions are conceivable. All music conditions in this study were played back via external loudspeakers.
%All music conditions in this study were played back via external loudspeakers. 
Although we show that non-stationary acoustic content such as music does not fundamentally limit the proposed approach, direct music playback via the earphones themselves introduces additional challenges, including acoustic feedback paths and self-generated sound. While the frequency characteristics of music are not inherently problematic for the method, these aspects were not explicitly evaluated in this work. However, addressing these aspects would constitute an additive capability rather than a prerequisite for respiration monitoring, given that headphones are frequently worn without active audio. 
Incorporating playback-aware signal separation would therefore primarily broaden the range of supported use cases rather than address a fundamental limitation of the proposed approach.
%Incorporating playback-aware signal separation would extend the applicability of the proposed approach to mixed sensing--playback scenarios.

\textbf{Application-driven directions.} 
From an application perspective, the presented system is particularly well suited for contexts in which earphones are already worn for sensory down-regulation or relaxation without music. 
%The presented system is particularly well suited for contexts in which earphones are already worn for sensory down-regulation or relaxation without music.
In such settings, continuous breathing-rate monitoring can serve as a lightweight physiological indicator of relaxation and mental state, opening opportunities for real-time feedback, guided breathing interventions, and long-term tracking of stress-related trends. Moreover, the observed error characteristics suggest potential benefits from confidence-aware fusion with complementary modalities (e.g., IMU- or PPG-based respiration), especially for handling transient outliers that cannot be fully addressed through channel discrepancy alone.

\textbf{Further extensions.} Beyond respiration rate estimation, the extracted respiration-related signals may enable the estimation of additional respiratory parameters, such as inspiration-expiration timing or tidal volume. While not evaluated in this work, these parameters represent a natural extension of the proposed pipeline and could further increase its physiological relevance.

Together, these directions highlight how the proposed pipeline (1) enables fully on-device respiration rate estimation on commodity earphones, (2) supports reliable monitoring in low-motion and relaxation-oriented contexts, and (3) provides a foundation for richer respiratory analytics and application-driven extensions on everyday earables.
\section{Conclusion}
\label{sec:conclusion}
In this paper, we introduced \sysName{}, a lightweight and robust system for on-device respiration rate estimation on commercial earphones. By leveraging the dual in-ear and outer-ear microphone configuration of ANC-enabled headphones, the system employs LMS-based adaptive noise suppression to attenuate ambient noise captured by the outer microphone while preserving respiration-related acoustic components in the in-ear signal.
In a comprehensive user study with \numPart{} participants conducted under 
%a range of 
realistic acoustic conditions, we demonstrated that adaptive noise suppression can effectively recover respiratory sounds even under high noise exposure exceeding 80\,dB SPL, where offline-trained, fixed-filter approaches reach their limits. Across all evaluated
%acoustic 
conditions, \sysName{} maintains consistently high respiration rate estimation accuracy, achieving a mean absolute error (MAE) of \maeFuse{}\,\cpm{} using binaural fusion. By incorporating automatic outlier rejection based on channel discrepancy, the MAE is further reduced to \maeConf{}\,\cpm{}. Moreover, an MAE as low as 0.14\,\cpm{} within a robust $3\sigma$-interval 
llustrates the potential of confidence-aware fusion with
%indicates that additional gains are achievable through confidence-aware fusion 
with complementary sensing modalities.
We further demonstrated that the entire processing pipeline can be executed in real time on ANC-capable earphones, imposing less than 2\% computational load on the system and without requiring any audio data to leave the device. This was validated through a full deployment of \sysName{} on the OpenEarable~2.0 platform.
Overall, \sysName{} 
addresses both the stringent resource constraints of wearable devices
%addresses both the stringent computational and memory constraints of wearable devices 
and the privacy concerns associated with continuous audio sensing. These results highlight the potential of adaptive signal processing as a viable foundation for robust, unobtrusive vital sign monitoring on everyday earphones and open avenues for future work on advanced respiratory analytics and health-oriented applications. Beyond respiration sensing, the proposed noise suppression approach could be extended to other audio-based physiological signals, such as heart sounds, enabling future applications in cardiovascular health monitoring.

\begin{acks}
This work was partially funded by the Deutsche Forschungsgemeinschaft (DFG, German Research Foundation) – GRK2739/1 - Project Nr. 447089431 – Research Training Group: KD2School – Designing Adaptive Systems for Economic
Decisions and by the Carl-Zeiss-Stiftung (Carl-Zeiss-Foundation) as part of the project "JuBot - Staying young with robots".
\end{acks}

\bibliographystyle{ACM-Reference-Format}
\bibliography{references.bib}

\newpage

\appendix

\section{Appendix}
\label{sec:appendix}

\begin{comment}
\subsection{Noise Suppression Algorithm}

\begin{algorithm}[h]
\caption{Delayed Leaky LMS Noise Suppression}
\label{alg:noise_suppression_vector}
\KwIn{$x[n]$ outer mic (reference), $d[n]$ inner mic (primary)}
\KwIn{delay $\Delta=\delay{}$, taps $M=\ntaps{}$, step size $\mu$, leakage $\gamma$, clip $\tau$}
\KwOut{$y[n]$ estimated noise, $e[n]$ noise-reduced signal}

\BlankLine
\textbf{// Align channels}\;
$\tilde{x}[n] \leftarrow x[n-\Delta]$\;
$\tilde{d}[n] \leftarrow d[n]$\tcp*{using the aligned index range}

\BlankLine
\textbf{// Initialize}\;
$\mathbf{w}_0 \leftarrow \mathbf{0} \in \mathbb{R}^{M}$\;
$\nu \leftarrow \mu\gamma$\;

\BlankLine
\For{$n = 0$ \KwTo $N-1$}{
  \textbf{// Construct reference vector}\;
  $\mathbf{x}_n \leftarrow [\tilde{x}[n],\tilde{x}[n-1],\ldots,\tilde{x}[n-M+1]]^\top$\;

  \BlankLine
  \textbf{// Estimate noise and subtract}\;
  $y[n] \leftarrow \mathbf{w}_n^\top \mathbf{x}_n$\;
  $e[n] \leftarrow \tilde{d}[n] - y[n]$\;

  \BlankLine
  \textbf{// Gradient norm and clipping scale (vector form)}\;
  $g_n \leftarrow e[n]\mathbf{x}_n$\tcp*{normalized gradient direction}
  $s_n \leftarrow \min\!\left(1,\dfrac{\tau}{\|g_n\|_2}\right)$\tcp*{clip factor}

  \BlankLine
  \textbf{// Leaky LMS update (with clipped gradient)}\;
  $\mathbf{w}_{n+1} \leftarrow (1-\nu)\mathbf{w}_n - \mu\, s_n\, g_n$\;
}
\Return{$e[n]$}\;
\end{algorithm}

\end{comment}

\subsection{Dataset}

\begin{table}[h]
\centering
\caption{Ground-truth respiration rate distribution in CPM by prior Exercise.}
\label{tab:activity}
\begin{tabular}{lcccccc}
\toprule
\textbf{Category} &
\textbf{Mean} &
\textbf{Median} &
\textbf{Std} &
\textbf{Min} &
\textbf{Max} &
\textbf{Range} \\
\midrule
No Exercise   & 16.82 & 16.53 & 3.77 & 7.57 & 26.24 & 18.67 \\
Post Exercise & 18.20 & 18.44 & 3.97 & 7.77 & 27.94 & 20.17 \\
\midrule
Overall       & 17.57 & 17.47 & 3.94 & 7.57 & 27.94 & 20.37 \\
\bottomrule
\end{tabular}
\end{table}

\subsection{Evaluation per Activity}
\begin{table}[h]
\centering
\caption{Breathing-rate estimation error by activity. MAE and RMSE are reported for single-channel, fused, and confident estimates. Values are given as mean $\pm$ standard deviation in CPM.}
\label{tab:rr_activity}
\begin{tabular}{lcccccc}
\toprule
 & \multicolumn{3}{c}{\textbf{MAE (CPM)}} & \multicolumn{3}{c}{\textbf{RMSE (CPM)}} \\
\cmidrule(lr){2-4} \cmidrule(lr){5-7}
\textbf{Activity} 
 & \textbf{Single} & \textbf{Fused} & \textbf{Confident}
 & \textbf{Single} & \textbf{Fused} & \textbf{Confident} \\
\midrule
No Exercise
 & $1.22 \pm 2.50$ & $1.10 \pm 2.13$ & $0.56 \pm 1.69$
 & $2.78 \pm 2.78$ & $2.39 \pm 2.39$ & $1.78 \pm 1.78$ \\
Post Exercise
 & $0.63 \pm 2.03$ & $0.61 \pm 1.87$ & $0.41 \pm 1.50$
 & $2.13 \pm 2.12$ & $1.96 \pm 1.96$ & $1.55 \pm 1.55$ \\
\midrule
\textbf{Overall}
 & $0.90 \pm 2.27$ & $0.84 \pm 2.00$ & $0.47 \pm 1.58$
 & $2.45 \pm 2.45$ & $2.16 \pm 2.17$ & $1.65 \pm 1.65$ \\
\bottomrule
\end{tabular}
\end{table}

\subsection{Evaluation per Condition}

%\subsection{Evaluation Noise Suppression}

% \begin{table}[t]
% \centering
% \caption{Breathing-rate estimation error by condition using fused estimates. Mean absolute error (MAE) and root mean square error (RMSE) are reported for ANS, BreathPro, and a filter-only baseline.}
% \label{tab:cond_mae_rmse_ref_compare_fused}
% \begin{tabular}{lcccccc}
% \toprule
%  & \multicolumn{3}{c}{\textbf{MAE (CPM)}} & \multicolumn{3}{c}{\textbf{RMSE (CPM)}} \\
% \cmidrule(lr){2-4} \cmidrule(lr){5-7}
% \textbf{Condition} &
% \textbf{ANS} & \textbf{BreathPro} & \textbf{Filter only} &
% \textbf{ANS} & \textbf{BreathPro} & \textbf{Filter only} \\
% \midrule
% Baseline (<35dB)      & 0.77 & 0.68 & 0.71 & 1.96 & 1.82 & 1.97 \\
% 50dB White Noise      & 0.24 & 0.21 & 0.29 & 0.75 & 0.70 & 0.76 \\
% 65dB White Noise      & 0.28 & 0.87 & 2.57 & 0.99 & 1.77 & 3.90 \\
% 80dB White Noise      & 0.84 & 2.71 & 4.60 & 1.91 & 3.89 & 5.51 \\
% 65dB Cafeteria        & 0.46 & 3.62 & 4.11 & 1.21 & 4.66 & 5.15 \\
% 80dB Musik            & 0.55 & 6.03 & 6.35 & 1.60 & 7.28 & 7.48 \\
% \midrule
% \textbf{Overall}      & 0.53 & 2.34 & 3.07 & 1.48 & 4.00 & 4.67 \\
% \bottomrule
% \end{tabular}
% \label{tab:ns_rr_performance}
% \end{table}

\begin{table}[h]
\centering
\caption{Breathing-rate estimation error by condition using fused estimates. MAE and RMSE are reported for ANS, nLMS, Breath Pro, and a  bandpass-filtered-only (BPF only) baseline.}
\label{tab:ns_rr_performance}
\begin{tabular}{lcccccccc}
\toprule
 & \multicolumn{4}{c}{\textbf{MAE (CPM)}} & \multicolumn{4}{c}{\textbf{RMSE (CPM)}} \\
\cmidrule(lr){2-5} \cmidrule(lr){6-9}
\textbf{Condition} &
\textbf{ANS} & \textbf{nLMS} & \textbf{BreathPro} & \textbf{BPF only} &
\textbf{ANS} & \textbf{nLMS} & \textbf{BreathPro} & \textbf{BPF only} \\
% \midrule
% Baseline (<35dB)      
%  & 0.75 & 0.74 & 0.74 & 0.84
%  & 2.14 & 2.12 & 2.11 & 2.34 \\

% 50dB White Noise      
%  & 0.41 & 0.41 & 0.42 & 0.57
%  & 1.50 & 1.48 & 1.40 & 1.76 \\

% 65dB White Noise      
%  & 0.50 & 0.67 & 1.24 & 2.74
%  & 1.63 & 1.94 & 2.47 & 4.01 \\

% 80dB White Noise      
%  & 1.33 & 1.61 & 3.32 & 4.15
%  & 2.71 & 3.01 & 4.63 & 5.15 \\

% 65dB Cafeteria        
%  & 0.93 & 1.21 & 4.05 & 4.32
%  & 2.11 & 2.49 & 5.14 & 5.35 \\

% 80dB Musik            
%  & 0.74 & 1.98 & 7.00 & 7.36
%  & 2.29 & 3.85 & 8.35 & 8.58 \\

% \midrule
% \textbf{Overall}      
%  & 0.77 & 1.09 & 2.75 & 3.26
%  & 2.09 & 2.59 & 4.62 & 5.01 \\
% \bottomrule
\midrule
Baseline (<35\,dB)
 & 0.85 & 0.86 & 0.88 & 0.94
 & 2.27 & 2.26 & 2.38 & 2.37 \\
50\,dB White Noise
 & 0.47 & 0.41 & 0.44 & 0.66
 & 1.56 & 1.30 & 1.38 & 1.92 \\
65\,dB White Noise
 & 0.55 & 0.66 & 1.35 & 2.75
 & 1.67 & 1.81 & 2.58 & 3.84 \\
80\,dB White Noise
 & 1.43 & 1.62 & 3.12 & 4.18
 & 2.84 & 2.99 & 4.30 & 5.15 \\
65\,dB Cafeteria
 & 0.95 & 1.33 & 3.90 & 4.25
 & 2.07 & 2.61 & 5.00 & 5.25 \\
80\,dB Music
 & 0.81 & 1.82 & 5.65 & 6.07
 & 2.30 & 3.54 & 6.92 & 7.25 \\
\midrule
\textbf{Overall}
 & \textbf{0.84} & \textbf{1.11} & \textbf{2.54} & \textbf{3.11}
 & \textbf{2.16} & \textbf{2.52} & \textbf{4.19} & \textbf{4.65} \\
\bottomrule
\end{tabular}
\end{table}

\newpage
\subsection{Evaluation per Participant}

\begin{table*}[h]
\centering
\caption{Participant-wise respiration rate estimation performance.
Mean absolute error (MAE) and root mean square error (RMSE)
are reported for single-channel, fused (mean L/R), and confident estimates.
Values in parentheses denote performance after outlier removal using a 3$\sigma$ MAD criterion (single-channel and fused).
All values are given in CPM.}
\label{tab:participant_metrics_conf}
\begin{tabular}{lcccccc}
\toprule
\textbf{Participant} &
\multicolumn{3}{c}{\textbf{MAE (CPM)}} &
\multicolumn{3}{c}{\textbf{RMSE (CPM)}} \\
\cmidrule(lr){2-4} \cmidrule(lr){5-7}
 & \textbf{Single (3$\sigma$)} & \textbf{Fused (3$\sigma$)} & \textbf{Conf.}
 & \textbf{Single (3$\sigma$)} & \textbf{Fused (3$\sigma$)} & \textbf{Conf.} \\
\midrule
P01 & 0.28 (0.11) & 0.22 (0.10) & 1.05 & 1.14 (0.14) & 0.66 (0.13) & 1.35 \\
P02 & 1.60 (0.20) & 1.55 (0.18) & 1.68 & 3.71 (0.26) & 3.16 (0.23) & 2.86 \\
P03 & 0.35 (0.15) & 0.34 (0.14) & 1.10 & 1.29 (0.19) & 1.24 (0.18) & 2.04 \\
P04 & 0.59 (0.11) & 0.58 (0.11) & 1.59 & 2.14 (0.15) & 1.66 (0.14) & 2.35 \\
P05 & 2.80 (0.19) & 2.51 (0.18) & 2.42 & 4.97 (0.28) & 4.36 (0.25) & 4.06 \\
P06 & 0.69 (0.13) & 0.68 (0.12) & 1.38 & 2.22 (0.17) & 1.90 (0.16) & 2.41 \\
P07 & 2.26 (0.20) & 1.93 (0.18) & 2.49 & 4.20 (0.27) & 3.55 (0.24) & 3.31 \\
P08 & 0.53 (0.14) & 0.51 (0.13) & 1.55 & 1.57 (0.19) & 1.21 (0.17) & 1.90 \\
P09 & 0.77 (0.23) & 0.76 (0.22) & 1.43 & 1.77 (0.30) & 1.74 (0.29) & 2.07 \\
P10 & 2.75 (1.91) & 2.44 (1.55) & 3.50 & 4.14 (2.75) & 3.73 (2.24) & 4.66 \\
P11 & 0.61 (0.12) & 0.58 (0.11) & 1.71 & 1.94 (0.16) & 1.60 (0.14) & 2.77 \\
P12 & 2.59 (1.21) & 2.23 (1.04) & 2.29 & 3.93 (1.88) & 3.26 (1.59) & 3.17 \\
P13 & 0.72 (0.11) & 0.71 (0.10) & 1.20 & 2.38 (0.14) & 1.88 (0.13) & 1.97 \\
P14 & 0.47 (0.12) & 0.47 (0.11) & 1.81 & 1.63 (0.15) & 1.47 (0.14) & 3.15 \\
P15 & 1.50 (0.19) & 1.29 (0.18) & 3.26 & 3.09 (0.25) & 2.68 (0.23) & 4.02 \\
P16 & 2.09 (0.37) & 2.04 (0.36) & 3.65 & 3.98 (0.54) & 3.61 (0.53) & 4.77 \\
P17 & 1.23 (0.26) & 1.21 (0.25) & 1.94 & 3.08 (0.35) & 2.84 (0.34) & 3.01 \\
P18 & 0.35 (0.14) & 0.34 (0.14) & 0.34 & 1.43 (0.17) & 1.43 (0.17) & 1.43 \\
\midrule
\textbf{Overall}
& 1.21 (0.16) & 1.10 (0.15) & 0.58
& 2.92 (0.21) & 2.51 (0.19) & 1.92 \\
\bottomrule
\end{tabular}
\end{table*}

\begin{figure}[!h]
    \centering
    \includegraphics[width=0.95\linewidth]{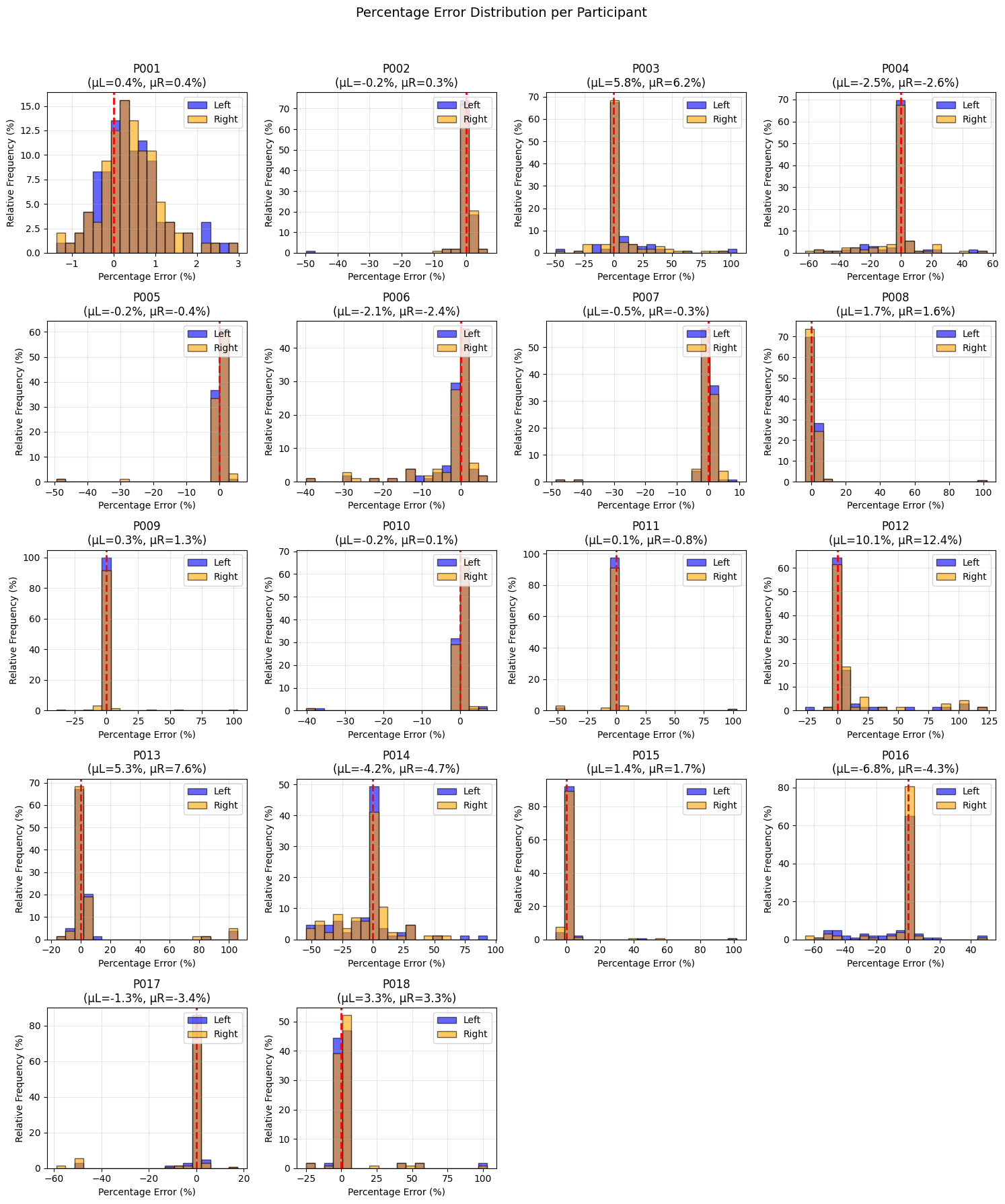}
    \caption{Distribution of percentage \RR{} error for each participant}
    \label{fig:participant-rr}
\end{figure}

\end{document}